\documentclass[3p,amsmath,amssymb,compress]{elsarticle}
\usepackage{amsmath}
\usepackage{graphicx}
\usepackage{latexsym}
\usepackage{amsmath,amssymb}
\usepackage{bm} 
\usepackage{xcolor}
\usepackage{stackrel}
\usepackage{accents}
\journal{Annals of Physics special issue "Eliashberg theory at 60"}

\numberwithin{equation}{section}
\DeclareMathOperator{\sgn}{sgn}
\newcommand{\tgl}{\tau_{\mathrm{GL}}}

\begin{document}
\begin{frontmatter}	

\title{Fluctuational Anomalous Hall and Nernst Effects in Superconductors}

\author{Songci Li}
\author{Alex Levchenko}
\address{Department of Physics, University of Wisconsin-Madison, Madison, Wisconsin 53706, USA}

\date{March 2, 2020}

\begin{abstract}
In this work we develop microscopic kinetic theory of the anomalous Hall and Nernst effects in superconductors induced by fluctuations in the vicinity of the critical transition temperature. The technical analysis is carried out within the Matsubara diagrammatic technique and linear response Kubo formula with disorder averaging and an analytical continuation. It is shown that asymmetric skew-scattering due to spin-orbit interaction gives rise to a new anomalous Hall conductance promoted by Maki-Thompson interference and density of states fluctuational effects. The side-jump mechanism of the anomalous Hall effect is present only in the density of states contributions. The anomalous Nernst effect is found due to a special nonlinear Aslamazov-Larkin term from the quantum-crossing of interacting fluctuations in a Hikami box with the skew-scattering process. In addition to calculations in the weak-coupling BCS model we also sketch an approach for the account of fluctuational transport effects in the strong-coupling limit. In particular we explore an example of how Keldysh version of the Eliashberg theory can in principle be constructed and estimate fluctuation-induced conductivity within conventional electron-phonon coupling scenario. These ideas may pave the way for further developments of strong coupling theories in applications to fluctuations in unconventional superconductors and anomalous transport responses in particular.   
\end{abstract}


\end{frontmatter}

\tableofcontents


\section{Introduction and motivation}\label{sec:intro}

The Hall effect in superconductors has been a subject of intensive studies over the years. One usually distinguishes transport regimes below and above the transition temperature into the superconducting state. Below the critical temperature $T_c$ in the mixed state of type-II superconductors the longitudinal ($\sigma_{xx}$) and Hall ($\sigma_{xy}$) conductivities are dominated by the vortex physics and the flux-flow. The generally accepted description of this transport regime was given in the original studies of Bardeen-Steven \cite{BS} and Nozi\`eres-Vinen \cite{NV} who showed that $\sigma_{xx}\approx \sigma_{n} (H_{c2}/H)$ and $\sigma_{xy}\approx (\Omega\tau)\sigma_{xx}$, where $\sigma_n$ is the normal-state conductivity, $\tau$ is the elastic lifetime with respect to impurity scattering, and $H_{c2}$ is the upper critical field. The energy scale $\Omega\approx\Delta^2/\varepsilon_F\ll\Delta$, where $\Delta$ is the superconducting energy gap and $\varepsilon_F$ is the Fermi energy, is related to the energy spacings of the Caroli-De Gennes-Matricon discrete electronic levels localized within the vortex core \cite{CdGM}. The model of the flux-flow Hall effect reflects an idea that the region inside the vortex core may be considered just as a normal metal with respect to its electronic properties.  
The microscopic calculations of flux-flow conductivities based on the quasiclassical nonequilibrium diagrammatic technique and kinetic equation that followed \cite{Kravtsov,LO-FluxFlow,Dorsey-FluxFlow,Kopnin-Lopatin,Feigelman-Skvortsov,Kopnin-Vinokur} essentially confirmed the basic assumptions. They also clarified importance of the pinning and the physics of dissipative processes in the vortex core related to electron-phonon relaxation as well as additional dissipation induced by the motion of a vortex via rare nonadiabatic Landau-Zener processes of electron excitations between the core levels.    

The physical mechanisms of the Hall effect observed above the transition are different and governed by the emergence of superconducting fluctuations (SF). At the onset of $T_c$ fluctuations of the superconducting order parameter form a new branch of charged collective excitations that contribute to the electromagnetic response. Three main mechanisms of fluctuation-induced transport have been identified. The direct contribution of the fluctuating Cooper pairs was proposed by Aslamazov-Larkin \cite{AL}. Almost simultaneously an additional interference mechanism was found by Maki \cite{Maki}. The apparent divergence of the corresponding fluctuation correction was termed anomalous and Thompson later showed \cite{Thompson} how pair-breaking scattering (or dephasing more generally) regularizes Maki's result. Lastly, a typically weaker density of states (DOS) mechanism was pointed out by Abraham-Redi-Woo \cite{Abrahams} who accounted for the redistribution of the quasi-particle states near the Fermi level by the long-lived fluctuations that tend to open an energy gap. Shortly after the zero-field calculations were completed, Fukuyama-Ebisawa-Tsuzuki \cite{FET} provided generalizations of the theory to fluctuational regimes in the external magnetic field. They revealed significant difference between Maki-Thompson (MT) and Aslamazov-Larkin (AL) fluctuational Hall conductivities and elucidated the crucial role of electron-hole asymmetry.  

The research topic of transversal transport effects in superconductors was rejuvenated by the discovery of high-$T_c$ materials when an anomalous behavior of the Hall coefficient was observed upon cooling from the normal state \cite{Clayhold} and giant Nernst signal was detected at the onset of $T_c$ \cite{Xu-Ong,Wang-Ong,Wang-Li-Ong}. The sign reversal of the Hall resistivity observed across $T_c$ had been seen even earlier in the conventional superconductors such as niobium (Nb) and vanadium (V) for example, however this feature was originally believed to be of extrinsic origin due to pinning or possible thermoelectric effects. The subsequent detailed measurements in copper-oxides revealed much deeper universality of this phenomenon as sign change of the Hall effect was detected to take place in both hole-doped systems, such as YBa$_2$Cu$_2$O$_{7-x}$ \cite{Matsuda}, and electron-doped compounds such as Nd$_{2-x}$Ce$_x$CuO$_4$ \cite{Hagen}. Many other materials showed the same signature feature \cite{LeBoeuf} including most recently van der Waals structures based on atomically thin superconducting Bi$_{2.1}$Sr$_{1.9}$CaCu$_2$O$_{8+x}$ \cite{Kim-BSCCO}. An anomaly in the Nernst effect has been also confirmed in different superconducting materials  by various groups, see review in Ref. \cite{Behnia-Review} and references therein. These results triggered the resurgence of interest in the topic of superconducting fluctuations as a possible cause for the observed features. Theoretical calculations were revisited in the context of both Hall \cite{Varlamov-Livanov,Ullah-Dorsey,Aronov-Hikami,Aronov-Rapoport,AHL,Michaeli-Hall} and Nernst \cite{USH,Kontani,Ussishkin,Reizer,Serbyn,Michaeli-Nernst,AL-Nernst,Hettinger} effects, and additional new mechanisms were proposed ranging from phase fluctuations \cite{Podolsky,Raghu} to topological vortex excitations and Berry phase \cite{Ao-Thouless,FGLV,Sumiyoshi-Fujkimoto}. 
     
In a parallel vein of developments finite polar Kerr effect (PKE) was observed below $T_c$ in a broad class of complex superconductors including strontium ruthenate Sr$_2$RuO$_4$ \cite{Kerr-SrRuO}, some cuprates including YBa$_2$Cu$_3$O$_{6+x}$ \cite{Kerr-YBCO} and La$_{1-x}$Bs$_x$CuO$_4$ \cite{Kerr-LBCO}, as well as heavy-fermion systems UPt$_3$ \cite{Kerr-UPt}, URu$_2$Si$_2$ \cite{Kerr-URuSi}, and PrOs$_4$Sb$_{12}$ \cite{Kerr-PrOsSb}. The angle of Kerr rotation $\theta_K$ can be related to the frequency dependent Hall conductivity $\sigma_{xy}$. Thus observability of finite Kerr effect without external magnetic field at minimum requires broken time reversal symmetry. For this reason Kerr rotation measurements are considered to be an extremely sensitive probe of unconventional superconducting states. In the model calculations the microscopic origin of the Kerr effect was quickly suggested to be linked to the same intrinsic and extrinsic mechanisms of the anomalous Hall effect (AHE) as known to occur in ferromagnets \cite{Nagaosa-RMP}. These are the anomalous velocity and spin-orbit band effects due to Karplus-Luttinger \cite{Karplus-Luttinger}, skew-scattering introduced by Smit \cite{Smit}, and peculiar quantum side-jump revealed by Berger \cite{Berger}. Thus the existing analytical treatments of PKE can be roughly split into two main groups: clean multiband models \cite{Mineev,Kallin,Gradhand} and single band disorder models \cite{Goryo,Lutchyn,Li,Konig}. Alternative scenarios for the possible explanations of the Kerr effect in superconductors have also been proposed including spin-fluctuation mechanism \cite{Chubukov-Kerr} and persistent loop current correlations \cite{Brydon}. 

The intent of this work is to bring together main concepts from the theories of superconducting fluctuations and anomalous Hall effects to investigate fluctuation-induced anomalous responses in superconductors. In our recent study \cite{SL-AL} we have demonstrated that interaction effects in the Cooper channel and the related quantum interference corrections can be important for the temperature dependence of the anomalous Hall conductivity even for nominally non-superconducting materials. In the present paper we tailor and expand our analysis specifically to superconducting systems in the proximity of $T_c$. 

The rest of the paper is organized as follows. In Sec. \ref{sec:technique} we briefly summarize main aspects of the diagrammatic technique in application to disordered superconducting systems and present extensions of the methods needed to capture anomalous transversal transport coefficients. In Sec. \ref{sec:sk} we discuss the skew-scattering mechanism of the anomalous Hall effect for both Maki-Thompson and density of states contributions. A similar analysis is carried out in Sec. \ref{sec:sj} for the side-jump mechanism.  In Sec. \ref{sec:nonlinear} we consider nonlinear fluctuation effects in the context of Aslamazov-Larkin contribution to fluctuation-induced transport. We introduce new element of the theory, skew-scattering Hikami box, that connects fluctuating modes and renders finite anomalous Nernst effect (ANE) provided particle-hole asymmetry. Finally in Sec. \ref{sec:eliashberg} we lay out main ideas for the extension of weak-coupling results of superconducting fluctuations to the domain of strong coupling by employing Eliashberg's theory. Sec. \ref{sec:summary} contains a snapshot of main results in a table format and an outlook discussion of remaining open interesting questions for the future research.         
 

\section{Technical approach, microscopic formalism, and assumptions }\label{sec:technique}

In this section we briefly review main ingredients and building blocks of the diagrammatic technique for disordered electronic systems, provide basic definitions, and introduce key notations. This material is well established and covered in multiple textbooks \cite{AGD,LV-Book,AM-Book}, nevertheless we choose to keep it here to have a self-contained presentation and to explain clearly essential generalizations needed in the context of AHE. Throughout the paper we work in the units $\hbar=k_B=1$. 

For electrons moving in the random potential whose correlation function is given by the Gaussian white-noise, self-energy of the single-particle Green's function reduces to a single graph as all other crossed-type and rainbow-type diagrams with impurity lines are suppressed in powers of the large parameter $\varepsilon_F\tau\gg1$ \cite{AGD}. The resummation of leading term in the Dyson equation generates simple geometric series that leads to a classical result for the disorder-averaged Green's function in the form 
\begin{equation}\label{eq:G}
G_{\bm{p}}(\varepsilon_n)=(i\bar{\varepsilon}_n-\xi_{\bm{p}})^{-1}, \quad \bar{\varepsilon}_n=\varepsilon_n+\frac{1}{2\tau}\sgn(\varepsilon_n), \quad \xi_{\bm{p}}=\frac{p^2}{2m}-\varepsilon_F
\end{equation}
where $\varepsilon_n=(2n+1)\pi T$ is the fermionic Matsubara frequency. In this model the impurity scattering time is given by $\tau^{-1}=2\pi\nu n_{\text{imp}}V^2_0$ to the leading order in the Born approximation for scattering probability, where $\nu$ is the density of states at the Fermi energy, $n_{\text{imp}}$ is the impurity concentration, and $V_0$ is the zero-momentum Fourier transform of the impurity potential $V(\bm{r})$. As local attractive  interactions in the BCS model lead to a formation of Cooper pairs another quantum coherent effect of disorder manifests in the vertex renormalization in the particle-particle channel as electrons forming a pair experience scattering on the same impurity. In general the two-particle vertex function $\lambda_{\bm{q}}(\varepsilon,\varepsilon')$ satisfies the integral Bethe-Salpeter equation, however for the disorder model under considerations it simplifies to an algebraic form. In complete analogy to the Green's function calculation, the resummation of ladder diagrams gives for the vertex function in the momentum prepresentation 
\begin{equation}
\lambda_{\bm{q}}(\varepsilon_n,\varepsilon_m)=1+\frac{1}{2\pi\nu\tau}\int_{\bm{p}}\lambda_{\bm{q}}(\varepsilon_n,\varepsilon_m)G_{\bm{p}+\bm{q}}(\varepsilon_n)G_{-\bm{p}}(\varepsilon_m).
\end{equation}
The resulting form of $\lambda_{\bm{q}}$ that follows from the solution of this equation strongly depends on the value of the parameter $T\tau$. 
In general one should distinguish three different fluctuational regimes in superconductors: the diffusive scattering $T\tau\ll1$, the ballistic limit $1\ll T\tau\ll\sqrt{T_c/(T-T_c)}$, and the ultra-clean limit $T\tau\gg\sqrt{T_c/(T-T_c)}$. We will primarily concentrate on the diffusive case, which is also mathematically simpler. In the ballistic case, fluctuation effects become strongly non-local in space which brings additional significant complications. In calculating the two-particle $GG$-block in the diffusive limit it is sufficient to expand the Green's function at small momentum $G_{\bm{p}+\bm{q}}\approx G_{\bm{p}}+(\bm{vq})G^2_{\bm{q}}+(\bm{vq})^2G^3_{\bm{p}}$, where $\bm{v}=\bm{p}/m$. Then upon momentum integration and angular average over the direction on the Fermi surface $\int_{\bm{p}}\to\nu\int d\xi_{\bm{p}}dO_{\bm{p}}/4\pi$ one arrives at 
\begin{equation}\label{eq:lambda}
\lambda_{\bm{q}}(\varepsilon_n,\varepsilon_m)=\frac{|\bar{\varepsilon}_n-\bar{\varepsilon}_m|}{|\varepsilon_n-\varepsilon_m|+Dq^2\Theta(-\varepsilon_n\varepsilon_m)}
\end{equation}
where $D$ is the diffusion coefficient of the conduction electrons and $\Theta(x)$ is the Heaviside step-function.  

In the weak-coupling BCS theory with attractive interaction constant $g$, electron interactions in the Cooper channel are characterized by the propagator $L_{\bm{q}}(\Omega_k)$. The latter is found from the Dyson equation by summing a sequence of $GG$-loop diagrams with disorder renormalizations in the ladder approximation. This gives 
\begin{equation}\label{eq:L-Dyson}
L^{-1}_{\bm{q}}(\Omega_k)=-g^{-1}+T\sum_{\varepsilon_n}\int_{\bm{p}}\lambda_{\bm{q}}(\varepsilon_n+\Omega_k,-\varepsilon_n)G_{\bm{p}+\bm{q}}(\varepsilon_n+\Omega_k)G_{-\bm{p}}(-\varepsilon_n).
\end{equation}  
With the explicit forms of $G_{\bm{p}}$ from Eq. \eqref{eq:G} and $\lambda_{\bm{q}}$ from Eq. \eqref{eq:lambda} the momentum integration followed by the frequency summation gives 
\begin{equation}\label{eq:L}
L^{-1}_{\bm{q}}(\Omega_k)=-\nu\left[\ln\frac{T}{T_c}+\psi\left(\frac{1}{2}+\frac{Dq^2+|\Omega_k|}{4\pi T}\right)-\psi\left(\frac{1}{2}\right)\right],
\end{equation}
where $\psi(x)$ is the digamma function, and the critical temperature $T_c=(2\gamma_E\omega_D/\pi)\exp[-1/(\nu g)]$ was expressed through the bare BCS coupling constant $g$, with $\gamma_E=1.78$ being the Euler constant and $\omega_D$ being the Debye frequency which cuts logarithmically divergent summation at $n_{\text{max}}=\omega_D/2\pi T$ in the $\lambda GG$-polarization operator. At temperatures close to $T_c$, $\ln(T/T_c)\approx (T-T_c)/T_c$, the pole structure of the propagator at $(q,\Omega)\to0$ suggests a simplified form, which is the most useful in practical calculations 
\begin{equation}\label{eq:L-approx}
L_{\bm{q}}(\Omega_k)\approx-\frac{8T_c}{\pi\nu}\frac{1}{\tgl^{-1}+Dq^2+|\Omega_k|}, \quad \tgl=\frac{\pi}{8(T-T_c)} ,
\end{equation}
where we introduced Ginzburg-Landau time $\tgl$ and used $\psi'(1/2)=\pi^2/2$ when expanding the digamma function. Thus bosonic fluctuation modes are soft with characteristic energies $(Dq^2,\Omega)\sim T-T_c$ which are much smaller than typical excitation energies of fermionic quasiparticles $\varepsilon\sim T$.  This fact brings technical advantages in the analysis of diagrams relevant for transport processes and strong coupling effects. 

In the preceding considerations we assumed that impurity scattering of electrons does not involve spin degree of freedom. There are two main sources of spin-related scattering of conduction electrons originating either from localized spins of magnetic impurities or spin-orbit (SO) interaction. The latter can be characterized by the scattering amplitude $\propto[\bm{\sigma}\cdot(\bm{p}\times\bm{p}')]$, where $\bm{p}$ and $\bm{p}'$ are the initial and final momenta of an electron, and $\bm{\sigma}$ is the spin operator whose components are the Pauli matrices. To account for the SO effects we will use the following momentum dependent scattering amplitude
\begin{equation}\label{eq:V}
V_{\bm{pp}'}=V_0\left[1-\frac{i\alpha_{\text{so}}}{p^2_F}(\bm{p}\times\bm{p}')_z\right],
\end{equation}
where $\alpha_{\text{so}}$ is the dimensionless coupling constant proportional to the average of the spin  $\langle\sigma_z\rangle$, which could be assigned either to polarization of conduction carriers or to the localized impurity. In the context of disordered superconductors this amplitude appeared perhaps for the first time in the work of Abrikosov-Gor'kov on the problem of the Knight shift \cite{AG-Knight}. In the context of transport properties it was employed by Hikami-Larkin-Nagaoka in their seminal work on the weak antilocalization \cite{HLN}. For our purposes, the form of this spin and momentum dependent scattering potential generates two important effects. The cross product between the first and the second term in Eq. \eqref{eq:V} leads to skewness in the differential scattering cross-section of electronic transport. It also generates an additional term in the matrix element of the velocity operator which is responsible for the side-jump processes. In addition, the square of the second term in Eq. \eqref{eq:V} averaged over the directions of momenta leads to the dephasing in the Cooper channel. The latter feature shifts critical temperature by Abrikosov-Gor'kov mechanism and also provides regularization for the anomalous Maki-Thompson term. The spin-flip term neglected in Eq. \eqref{eq:V} gives the same effect on $T_c$ suppression and MT diagram. The dephasing time corresponding to $V_{\bm{pp}'}$ of Eq. \eqref{eq:V} is $\tau^{-1}_\phi=2\pi\nu n_{\text{imp}}\alpha^2_{\text{so}}V^2_0$, which shifts the diffusive term in the vertex function $Dq^2\to Dq^2+\tau^{-1}_\phi$ of Eq. \eqref{eq:lambda}, and a similar shift carries over to the pair propagator in Eq. \eqref{eq:L}. As a result, the pole of $L_{\bm{q}}(\Omega_k)$ in $(q,\Omega)\to0$ limit occurs at different $T_c$ 
\begin{equation}\label{eq:Tc}
\ln\left(\frac{T_c}{T_{c0}}\right)=\psi\left(\frac{1}{2}\right)-\psi\left(\frac{1}{2}+\frac{\Gamma}{4\pi T_c}\right).
\end{equation}
This equation implicitly defines the renormalized critical temperature $T_c$ as a function of the depairing energy scale $\Gamma=\tau^{-1}_\phi$.  A full analytic solution of this equation in terms of $T_c(\Gamma)$ is not possible, but asymptotic expressions can be easily extracted. For $\Gamma\ll T_{c0}$ we can expand the digamma function $\psi(x + 1/2)$ to first order in $x$ and thus obtain $\ln(T_{c0}/T_c)-\pi^2\Gamma/8\pi T_c=0$. To the first order in $\Gamma$ this yields $T_c\approx T_{c0}-\pi\Gamma/8$. The expansion for large values of $\Gamma$ is slightly more involved, because there exists a critical value of the pair-breaking parameter $\Gamma_c$ at which the critical temperature vanishes nonanalytically as a function of $\Gamma$. In order to see this, we rewrite Eq. \eqref{eq:Tc} in the following form $T_{c0}/T_c=\exp[\psi(1/2+\Gamma/4\pi T_c)]\exp[-\psi(1/2)]$. For the case of sufficiently strong depairing at which $T_c$ tends to vanish, we can make use of the asymptotic expansion $\exp[\psi(x + 1/2)]\approx x+1/(24x) +\ldots$ valid for large $x\gg1$. This expansion yields
$T_{c0}/T_c\approx4\gamma_E(\Gamma/4\pi T_c+\pi T_c/6\Gamma)$. The critical pair-breaking parameter is defined as the value at which the critical temperature vanishes $T_c(\Gamma_c)=0$. In this case we can neglect the second term on the right hand side of the last equation and obtain $\Gamma_c=\pi T_{c0}/\gamma_E\approx 1.76T_{c0}$. Expressed in terms of this quantity the critical temperature for $\Gamma\to\Gamma_c$ then behaves asymptotically as $T_c\approx(\sqrt{6}/2\pi)\sqrt{\Gamma_c(\Gamma_c-\Gamma)}$. In what follows we will focus our attention to the case of weak depairing $\Gamma\ll T_{c0}$. 


\section{Skew-scattering mechanism of the AHE in superconductors}\label{sec:sk}

We proceed to calculate the static conductivity, which in the linear response of Kubo formalism can be expressed in a standard way in terms of the operator of the electromagnetic response $Q_{ij}(\omega_\nu)$. It is calculated from the current-current correlation function that is expressed at Matsubara frequencies first and then an analytical continuation into the upper half-plane of complex frequency, $\omega_\nu\to-i\omega$, is needed to find the corresponding retarded kernel $Q^R_{ij}(\omega)$. In the dc-limit, $\omega\to0$, conductivity is given by $\sigma_{ij}=\lim_{\omega\to0}Q^R_{ij}(\omega)/(-i\omega)$. To find Hall conductivity we need to calculate the transversal part of $Q^R_{xy}$. 

\begin{figure}
\centering
\includegraphics[width=0.75\linewidth]{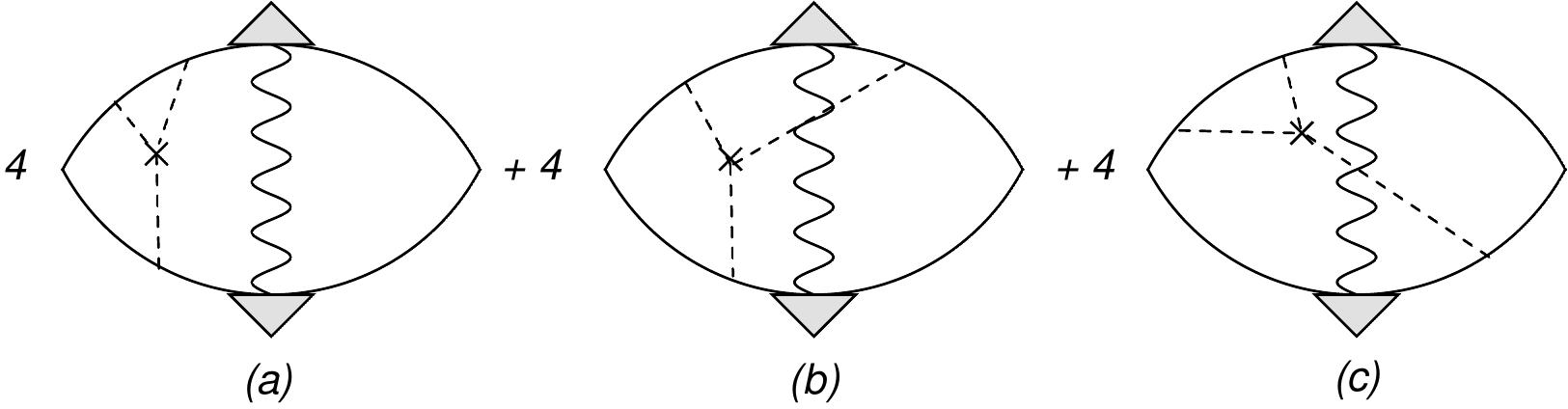}
\caption{Diagrams for the Maki-Thompson corrections to the AHE in the skew-scattering mechanism. The wavy line represents pair-propagator Eq. \eqref{eq:L}. The shaded triangles stand for the impurity ladder vertex function Eq. \eqref{eq:lambda}. The dashed lines represent scattering by an impurity, marked by a cross, with the corresponding potential from Eq. \eqref{eq:V}.  A factor of four near each diagram represents the total number of equivalent ways to arrange for impurity lines between Green's function per Eq. \eqref{eq:G} depicted by solid lines. }
\label{fig:SK-MT}
\end{figure}

\subsection{Maki-Thompson contribution to AHE}

We begin with the Maki-Thompson diagrams. There are three distinct possibilities shown in Fig. \ref{fig:SK-MT} each of which comes in four copies. We carefully remark 
that there are additional two diagrams containing an impurity vertex ladder that connects upper and lower Green's functions and crosses the line of a pair-propagator, the so-called cross Cooperon diagrams. As is known these terms are not important near $T_c$ in terms of their temperature dependence in $T-T_c$. In addition, in accordance with the established terminology, MT contribution is split into the regular and anomalous parts. In that context, the term anomalous references to the fact that that particular contribution is formally logarithmically divergent in two-dimensional case unless regularization due to pair breaking is included. We will focus on anomalous MT term as it carries the most singular temperature dependence.  

An analytical expression for the current-current response function corresponding to diagram-(a) in Fig. \ref{fig:SK-MT} can be written as  
\begin{equation}\label{eq:Q-sk-MT}
Q^{\text{sk-MT-a}}_{xy}(\omega_\nu)=2e^2T\sum_{\Omega_k}\int_{\bm{q}}  L_{\bm{q}}(\Omega_k)\Sigma^{\text{sk-MT-a}}_{\bm{q}}(\Omega_k,\omega_\nu).
\end{equation}
The internal block of this formula represents what one gets after the summation over the Fermionic momenta and frequencies, namely 
\begin{equation}
\Sigma^{\text{sk-MT-a}}_{\bm{q}}=T\sum_{\varepsilon_n}\lambda_{\bm{q}}(\varepsilon_n,\Omega_{k-n})
\lambda_{\bm{q}}(\varepsilon_{n+\nu},\Omega_{k-n-\nu})J^{\text{sk-MT-a}}_{xy},
\end{equation} 
where $\varepsilon_{n+\nu}\equiv\varepsilon_n+\omega_\nu, \Omega_{k-n}\equiv\Omega_k-\varepsilon_n$, and we assumed $\omega_\nu>0$ without loss of generality. The transversal current block
\begin{align}
&J^{\text{sk-MT-a}}_{xy}=n_{\text{imp}}\nu^3\langle v_{\bm{p},x} v_{-\bm{k},y}V_{\bm{p}\bm{k}'}V_{\bm{k}'\bm{k}}V_{\bm{k}\bm{p}} \rangle\nonumber \\ 
&\times\int d\xi_{\bm{k}} G_{\bm{k}}(\varepsilon_n)G_{\bm{k}}(\varepsilon_{n+\nu})G_{-\bm{k}}(\Omega_{k-n-\nu})G_{-\bm{k}}(\Omega_{k-n}) 
 \int d\xi_{\bm{p}} d\xi_{\bm{k}'} G_{\bm{p}}(\varepsilon_n)G_{\bm{p}}(\varepsilon_{n+\nu})G_{\bm{k}'}(\varepsilon_{n+\nu})
\end{align}
contains an angular average $\langle\ldots\rangle$ over the directions of momenta. By carrying out integrals over the fermionic dispersions, performing angular average over the Fermi surface followed by the Fermionic frequency summation, we obtain
\begin{align}
&\Sigma^{\text{sk-MT-a}}_{\bm{q}}(\Omega_k \geq 0,\omega_\nu) = \frac{2\pi^2D}{3}n_{\text{imp}}\alpha_{\text{so}}\nu^3V^3_0\tau \frac{\Theta(\omega_\nu-\Omega_{k+1})}{\omega_\nu+Dq^2+\Gamma} \nonumber \\
	 &\times \Bigg[\psi\left(\frac{1}{2}+\frac{2\omega_\nu-\Omega_k+Dq^2+\Gamma}{4\pi T}\right)-\psi\left(\frac{1}{2}+\frac{\Omega_k+Dq^2+\Gamma}{4\pi T}\right)\Bigg].
\end{align}
The step-function here implies that the summation over $\Omega_k$ is confined within the external frequency $\omega_\nu$. The remaining Matsubara sum over the bosonic frequencies $\Omega_k$ in Eq. \eqref{eq:Q-sk-MT} is done via the contour integral in the complex plane, so that after the analytical continuation we arrive at conductivity  
\begin{equation}
\sigma^{\text{sk-MT-a}}_{xy}=\frac{8\pi^2}{3}\sigma_Q(T\tau)\frac{D}{\tau_{\text{sk}}}\int_{\bm{q}}\frac{1}{(Dq^2+\tau^{-1}_\phi)(Dq^2+\tgl^{-1})}
\end{equation}
where we have used Eqs. \eqref{eq:lambda} and \eqref{eq:L-approx} and introduced the characteristic skew-scattering time $\tau^{-1}_{\text{sk}}=n_{\text{imp}}\alpha_{\text{so}}\nu^2V^3_0$. The notation $\sigma_Q=e^2/(2\pi\hbar)$ is for the quantum of conductance. The evaluation of the remaining diagrams in Fig.\ref{fig:SK-MT} reveals that $\sigma^{\text{sk-MT-b}}_{xy}=-\sigma^{\text{sk-MT-a}}_{xy}$ and $\sigma^{\text{sk-MT-c}}_{xy}=(1/2)\sigma^{\text{sk-MT-a}}_{xy}$. Therefore
\begin{equation}\label{eq:sigma-sk-MT}
\sigma^{\text{sk-MT}}_{xy}=4\left[\sigma^{\text{sk-MT-a}}_{xy}+\sigma^{\text{sk-MT-b}}_{xy}+\sigma^{\text{sk-MT-c}}_{xy}\right]=\frac{4\pi}{3}\sigma_Q(\tau/\tau_{\text{sk}}) (T_c\tgl)\ln(\tau_\phi/\tgl),
\end{equation}
where the final expression after momentum integration was specified to a two-dimensional case of superconducting fluctuations relevant to thin films whose thickness is small compared to Ginzburg-Landau coherence length. We have also assumed $\tau_\phi\gg\tgl$. It is worth noting that in terms of the main singularity in the temperature dependence, transversal MT conductivity in the skew scattering mechanism has the same asymptotic behavior as MT correction to the longitudinal conductivity, namely $\sigma^{\text{sk-MT}}_{xy}(T)\propto \sigma^{\text{MT}}_{xx}(T)$.    

\begin{figure}
\centering
\includegraphics[width=0.75\linewidth]{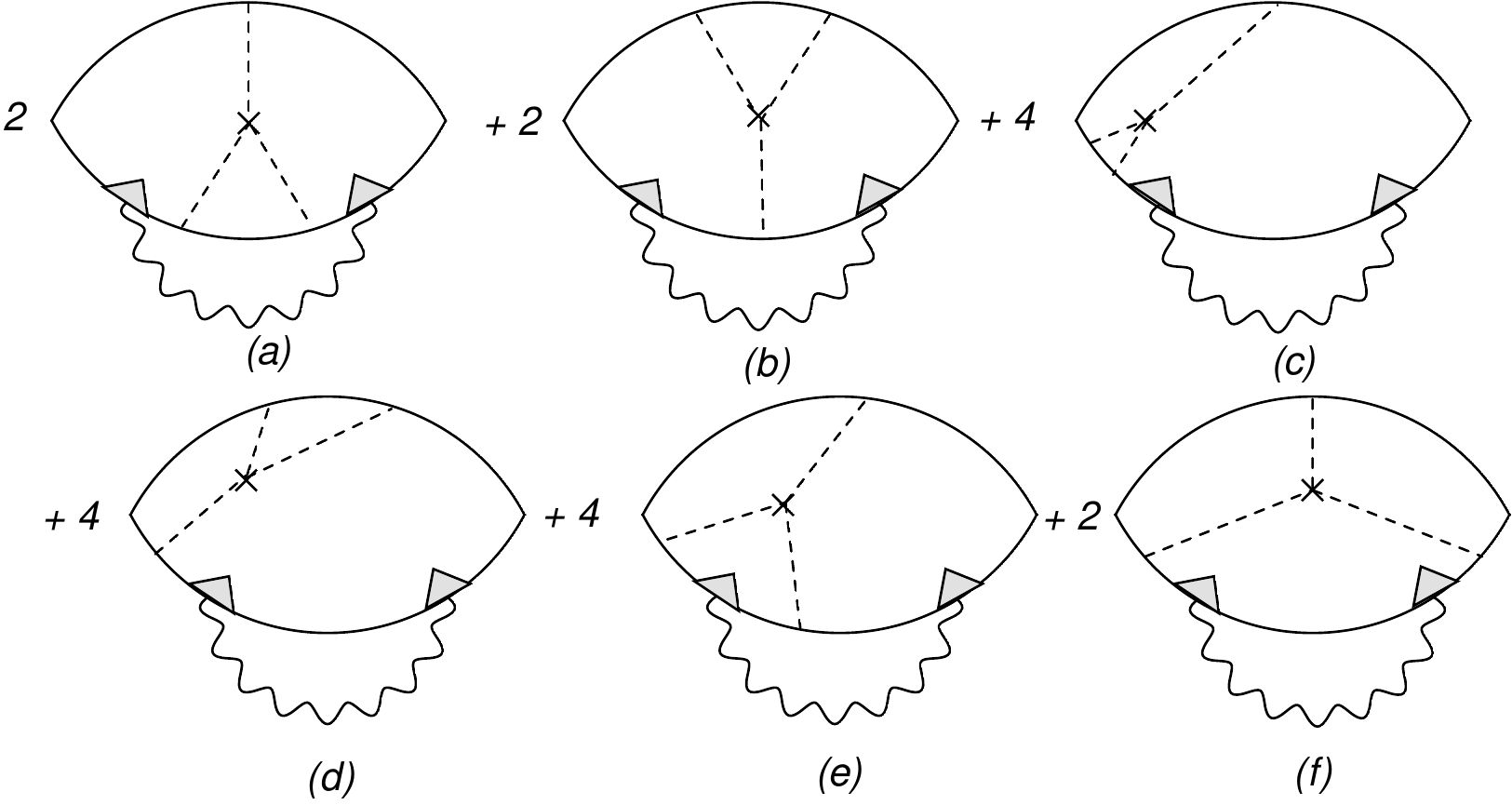}
\caption{Diagrams for the density of states corrections to the anomalous Hall conductivity in the skew-scattering mechanism. The convention for the diagrammatic blocks is the same as in Fig. \ref{fig:SK-MT}. The numerical coefficients in front of each diagram indicates the total number of possible rearrangements of skew impurity lines and fluctuation propagator line that lead to a similar contributions. }
\label{fig:SK-DOS}
\end{figure}

\subsection{Density of states contribution to AHE}

We continue with the analysis of the density of states diagrams, which are drawn in Fig. \ref{fig:SK-DOS}. The structural form of the electromagnetic response kernel $Q^{\text{sk-DOS}}_{xy}$ corresponding to these diagrams is identical to that of Eq. \eqref{eq:Q-sk-MT} the only difference is in the self-energy block that now reads as follows for the (a)-diagram as an example
\begin{equation}
\Sigma^{\text{sk-DOS-a}}_{\bm{q}}=T\sum_{\varepsilon_n}\lambda^2_{\bm{q}}(\varepsilon_n,\Omega_{k-n})J^{\text{sk-DOS-a}}_{xy},
\end{equation}    
where 
\begin{align}
&J^{\text{sk-DOS-a}}_{xy}=n_{\text{imp}}\nu^3
\langle v_{\bm{p},x} v_{\bm{k},y}V_{\bm{p}\bm{k}}V_{-\bm{p}-\bm{k}'}V_{-\bm{k}'-\bm{k}}\rangle\nonumber \\ 
&\times\int d\xi_{\bm{k}'}G_{-\bm{k}'}(\varepsilon_{n+\nu})
\Big[\int d\xi_{\bm{k}} G_{-\bm{k}}(\varepsilon_{n+\nu})G_{\bm{k}}(\Omega_{k-n})G_{\bm{k}}(\Omega_{k-n-\nu})\Big]^2.
\end{align}
After carrying out integrals over fermionic dispersions followed by the angular averages over the Fermi surface, we get
\begin{eqnarray}
	\Sigma^{\text{sk-DOS-a}}_{\bm{q}} =\frac{4\pi^3DT}{3}n_{\text{imp}}\alpha_{\text{so}}\nu^3V^3_0\tau \left(\sum_{\varepsilon_n>0}-\sum_{\varepsilon_n<0}+2\sum_{-\omega_\nu<\varepsilon_n<0}\right) \frac{\Theta(-\varepsilon_n\Omega_{k-n})}{\left(|2\varepsilon_n-\Omega_k|+Dq^2\right)^2}, 
\end{eqnarray}
where at the intermediate step we have splitted the summation in the frequency interval $(-\infty,-\omega_\nu)$ into the domains of $\varepsilon_n<0$ and $-\omega_\nu<\varepsilon_n<0$. 
Also in the denominator of the Cooperon we neglected depairing term $\Gamma$ as it does not play the same significant role as in the previous case of MT contributions. The summations in the frequency interval $\varepsilon_n>0$ and $\varepsilon_n<0$ are independent of the external frequency $\omega_\nu$ and are cancelled out by their mirror image diagrams. Keeping only the $\omega_\nu$ dependent term, we have
\begin{align}
\Sigma^{\text{sk-DOS-a}}_{\mathbf{q}}(\Omega_k,\omega_\nu)
=\frac{\pi \nu D}{6T}(\tau/\tau_{\text{sk}}) \Bigg\{\Theta(\Omega_k)\left[\psi'\left(\frac{1}{2}+\frac{2\omega_\nu+|\Omega_k|+Dq^2}{4\pi T}\right)-\psi'\left(\frac{1}{2}+\frac{|\Omega_k|+Dq^2}{4\pi T}\right)\right] \nonumber \\
+ \Theta(-\Omega_k)\Theta(\Omega_k+\omega_{\nu-1})\left[\psi'\left(\frac{1}{2}+\frac{2\omega_\nu-|\Omega_k|+Dq^2}{4\pi T}\right)-\psi'\left(\frac{1}{2}+\frac{|\Omega_k|+Dq^2}{4\pi T}\right)\right]\Bigg\}.
\end{align}
As the next step we substitute this $\Sigma^{\text{sk-DOS-a}}_{\mathbf{q}}$ into the response function $Q^{\text{sk-DOS}}_{xy}$, carry out the Matsubara sum over $\Omega_k$, perform analytical continuation, and expand the result to the linear order in external frequency. These steps lead us to the corresponding conductivity    
\begin{equation}
\sigma^{\text{sk-DOS-a}}_{xy}=-\frac{56\zeta(3)}{3\pi}\sigma_Q\frac{\tau D}{\tau_{\text{sk}}}\int_{\bm{q}}\frac{1}{Dq^2+\tgl^{-1}}
\end{equation}
where $\zeta(x)$ is Riemann zeta-function. In the two-dimensional case, the remaining momentum integral is logarithmically divergent at the upper limit which should be cut at the scale $Dq^2\sim T$ since in deriving the above equation we used approximate form of the pair propagator Eq. \eqref{eq:L-approx} obtained after the expansion of the digamma function in small ratio $Dq^2/4\pi T\ll1$. This spurious divergence is a well known artifact in the computation of the DOS term. A more accurate regularization procedure changes only the numerical factor under the logarithm which is beyond the accuracy of our analysis. Evaluations of the remaining diagrams in Fig.~\ref{fig:SK-DOS} show that $\sigma^{\text{sk-DOS-b}}=\sigma^{\text{sk-DOS-a}}, \sigma^{\text{sk-DOS-c}}+\sigma^{\text{sk-DOS-d}}=\sigma^{\text{sk-DOS-a}}+\sigma^{\text{sk-DOS-b}}$, and $\sigma^{\text{sk-DOS-e}}=\sigma^{\text{sk-DOS-f}}=-\sigma^{\text{sk-DOS-a}}$. Therefore, in two dimensional case the total correction is 
\begin{align}\label{sk-DOS-total}
\sigma^{\text{sk-DOS}}_{xy}= 2\Big[\sigma^{\text{sk-DOS-a}}+\sigma^{\text{sk-DOS-b}}+2\sigma^{\text{sk-DOS-c}}+2\sigma^{\text{sk-DOS-d}}+2\sigma^{\text{sk-DOS-e}}+\sigma^{\text{sk-DOS-f}}\Big] \nonumber \\ =-\frac{28\zeta(3)}{\pi^2}\sigma_Q (\tau/\tau_{\text{sk}})\ln(T_c\tgl).
\end{align}
Similarly to the MT contribution transversal Hall DOS term has the same temperature dependence as the longitudinal DOS fluctuation correction to diagonal conductivity. However the negative sign of DOS terms as compared to MT interference has to do with the depletion of quasiparticle states near the Fermi level as superconducting fluctuations tend to open an energy gap. 


\section{Side-jump mechanism of the AHE in superconductors}\label{sec:sj}

The side jump effect is manifest in an additional term in the matrix element of the velocity operator due to spin-orbit coupling
\begin{equation}\label{eq:v-sj}
	\langle \bm{p}'|\hat{\bm{v}}| \bm{p}\rangle\!=\!\frac{\bm{p}}{m}\delta_{\bm{p}\bm{p}'}-\frac{i\alpha_{\text{so}}}{2m\varepsilon_F}\!\sum_j 
	V_{\bm{p}-\bm{p}'}e^{i(\bm{p}-\bm{p}')\cdot\bm{R}_j}[\bm{e_z}\times(\bm{p}-\bm{p}')]
\end{equation}
where $\bm{R}_j$ is the radius vector of a given impurity and $\bm{e_z}$ is the unit vector along $z$-axis. As a consequence of that, one has to retain this new term in the current vertex of the electromagnetic response kernel $Q_{xy}$. Since $Q_{xy}$ contains the product of two currents the cross term between the usual velocity and an anomalous velocity generates additional contributions in both MT and DOS terms. It is found that the combined effect of side-jump processes cancels in the MT interference channel but survives and yields finite result in the DOS terms. Below we present technical details that lead to these conclusions.    

\begin{figure}
\centering
\includegraphics[width=0.5\linewidth]{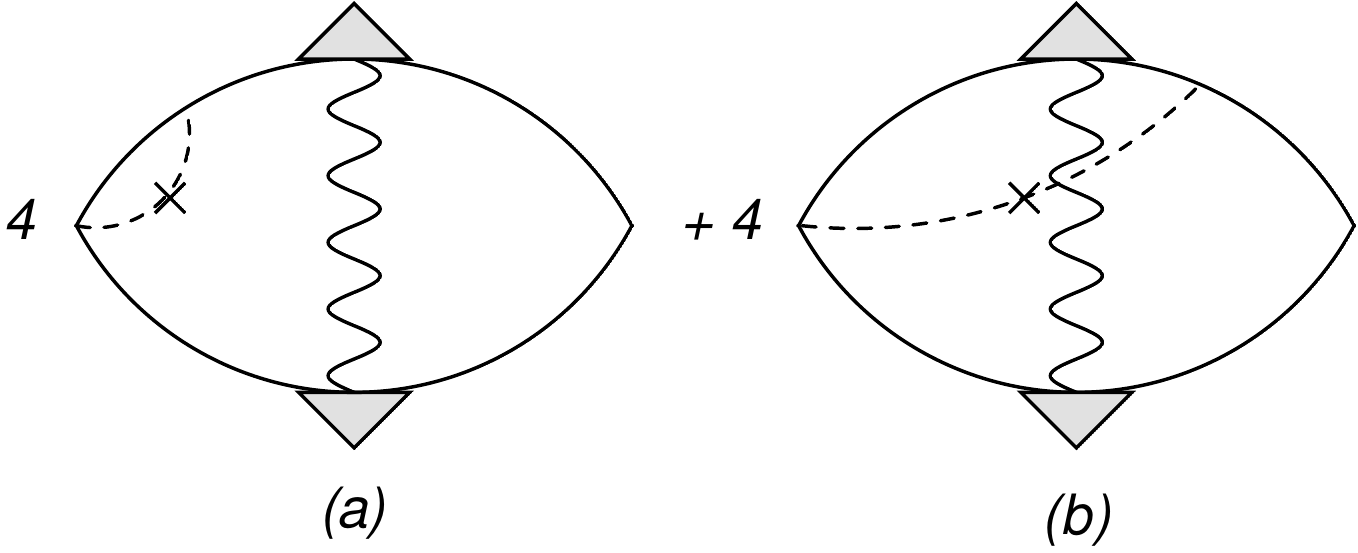}
\caption{Diagrams for the MT interference corrections in the side-jump mechanism. A factor of four in each case accounts for additional possibilities to connect impurity line to a different Green's function to extend this line from the different current vertex. }
\label{fig:SJ-MT}
\end{figure}

\subsection{Maki-Thompson contribution to AHE}

Diagrammatically a side-jump event corresponds to a particular diagrams when an impurity line extends from the current vertex. Physically it corresponds to the transversal displacement of the electronic wave-packet accumulated in successive scattering events. Two representative examples of side jumps in MT interference process are depicted in Fig. \ref{fig:SJ-MT}.   
For the diagram-(a) the respective self-energy block of the electromagnetic response reads 
\begin{equation}
\Sigma^{\text{sj-MT-a}}_{\bm{q}}(\Omega_k,\omega_\nu)=T\sum_{\varepsilon_n}\lambda_{\bm{q}}(\varepsilon_n,\Omega_k-\varepsilon_n)\lambda_{\bm{q}}(\varepsilon_{n+\nu},\Omega_k-\varepsilon_{n+\nu})J^{\text{sj-MT-a}}_{xy}
\end{equation}
where 
\begin{align}
&J^{\text{sj-MT-a}}_{xy}=-in_{\text{imp}}\nu^2\frac{\alpha_{\text{so}}V^2_0}{2m\varepsilon_F}\langle[\bm{e_z}\times(\bm{p}-\bm{p}')]_xv_{-\bm{p},y}\rangle\nonumber \\ 
&\times\int d\xi_{\bm{p}}d\xi_{\bm{p}'}G_{\bm{p}}(\varepsilon_n)G_{\bm{p}}(\varepsilon_{n+\nu})G_{-\bm{p}}(\Omega_k-\varepsilon_{n+\nu})G_{-\bm{p}}(\Omega_k-\varepsilon_n)G_{\bm{p}'}(\varepsilon_{n+\nu}).
\end{align}
After carrying out integrations over fermionic dispersions, we have
\begin{align}
	&\Sigma^{\text{sj-MT-a}}_{\bm{q}}(\Omega_k,\omega_\nu) = 4\pi^2 \nu^2\tau n_{\text{imp}}\frac{\alpha_{\text{so}}V^2_0}{2m\varepsilon_F}\langle[\bm{e_z}\times(\bm{p}-\bm{p}')]_xv_{-\bm{p},y}\rangle\nonumber \\
	& \times \left(\sum_{\varepsilon_n>0}-\sum_{\varepsilon_n<-\omega_\nu}+\sum_{-\omega_\nu<\varepsilon_n<0}\right) \frac{\Theta(-\varepsilon_n\Omega_{k-n})}{|2\varepsilon_n-\Omega_k|+Dq^2}\frac{\Theta(-\varepsilon_{n+\nu}\Omega_{k-n-\nu})}{|2\varepsilon_{n+\nu}-\Omega_k|+Dq^2}.
\end{align}
It can be shown following exactly the same steps of derivation that diagram-(b) in Fig. \eqref{fig:SJ-MT} gives the opposite contribution $\Sigma^{\text{sj-MT-b}}_{\bm{q}}=-\Sigma^{\text{sj-MT-a}}_{\bm{q}}$. 
Therefore 
\begin{equation}
	\sigma^{\text{sj-MT}}_{xy}=4\left(\sigma^{\text{sj-MT-a}}+\sigma^{\text{sj-MT-b}}\right)=0
\end{equation}
and we find complete cancellation in the net effect of side-jump processes in the MT channel.  

\begin{figure}
\centering
\includegraphics[width=0.5\linewidth]{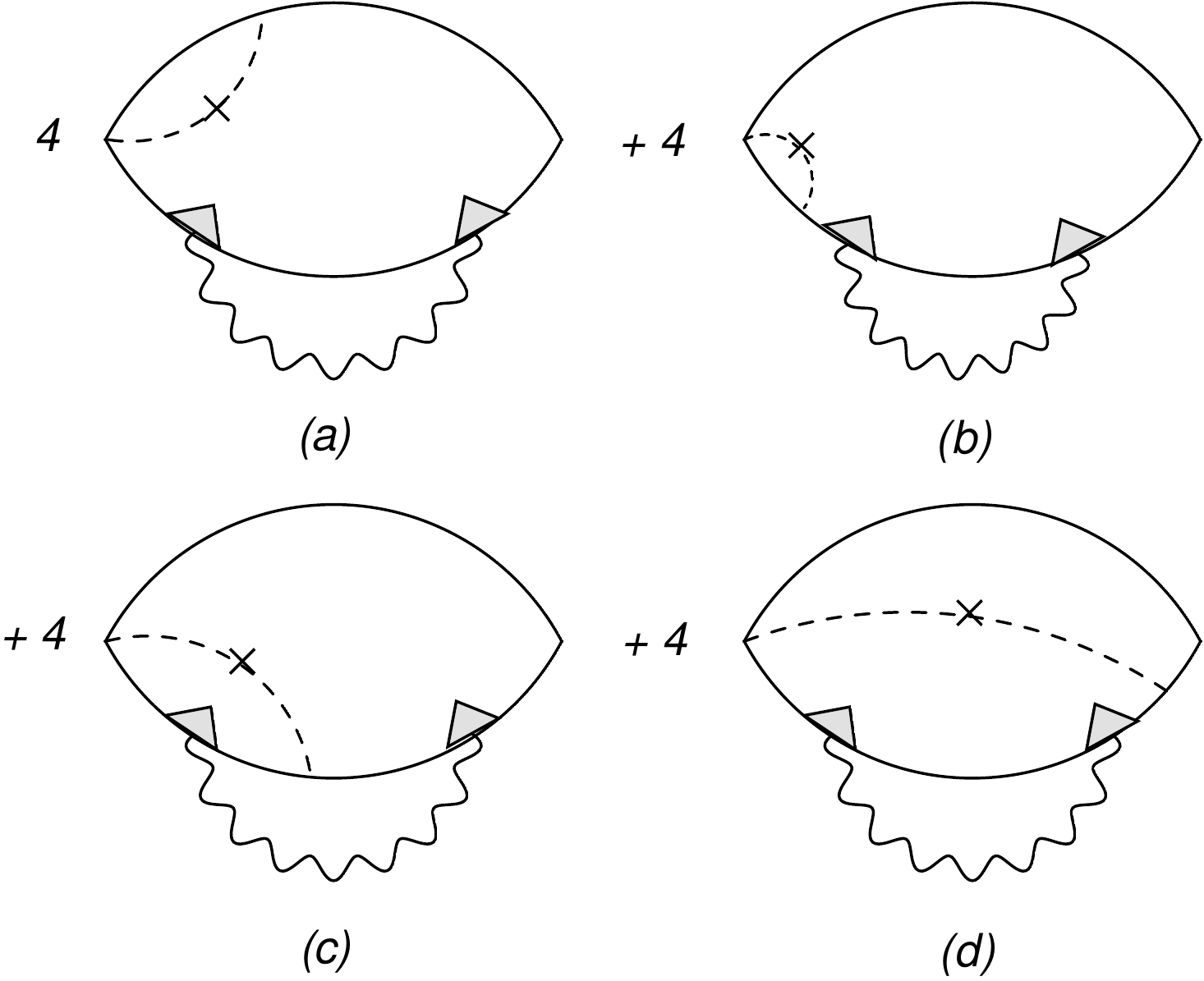}
\caption{Diagrams for the DOS fluctuational corrections in the side-jump mechanism. The conventions are the same as in Fig. \ref{fig:SJ-MT} with the same nomenclature for the combinatorial coefficients.}
\label{fig:SJ-DOS}
\end{figure}

\subsection{Density of states contribution to AHE}

The remaining terms to be considered are the side-jumps in the density of states fluctuations. The corresponding diagrams are displayed in Fig. \ref{fig:SJ-DOS}. The sum of self-energies for the first two diagrams is 
\begin{align}
\Sigma^{\text{sj-DOS-(a+b)}}_{\bm{q}}=T\sum_{\varepsilon_n}\lambda^2_{\bm{q}}(\varepsilon_n,\Omega_{k-n})J^{\text{sj-DOS-(a+b)}}_{xy},
\end{align}   
where 
\begin{align}
&J^{\text{sj-DOS-(a+b)}}_{xy}=-in_{\text{imp}}\nu^2\frac{\alpha_{\mathrm{so}}V^2_0}{2m\varepsilon_F}\langle[\bm{e_z}\times(\bm{p}-\bm{p}')]_xv_{\bm{p},y}\rangle\nonumber \\ 
&\times\int d\xi_{\bm{p}}\xi_{\bm{p}'} G^2_{\bm{p}}(\varepsilon_n)G_{\bm{p}}(\varepsilon_{n+\nu})G_{\bm{p}}(\Omega_{k-n})\Big[G_{\bm{p}'}(\varepsilon_{n+\nu})+ G_{\bm{p}'}(\varepsilon_n) \Big].
\end{align}
The momentum integrations and angular average in the current block $J_{xy}$ reduces self-energy to the following expression 
\begin{equation}
\Sigma^{\text{sj-DOS-(a+b)}}_{\bm{q}}=\frac{2\pi n_{\text{imp}}\alpha_{\text{so}}\nu V^2_0}{\varepsilon_F}(\pi \nu DT)\left(\sum_{\varepsilon_n>0}-\sum_{\varepsilon_n<-\omega_\nu}\right) \frac{\Theta(-\varepsilon_n\Omega_{k-n})}{(|2\varepsilon_n-\Omega_k|+Dq^2)^2}.
\end{equation}  
As in the previous case, it is technically advantageous to split the summation in the frequency interval $(-\infty,-\omega_\nu)$ into two regions 
$ \sum_{\varepsilon_n<-\omega_\nu}(\cdots)=\left(\sum_{\varepsilon_n<0}-\sum_{-\omega_\nu<\varepsilon_n<0}\right)(\cdots)$. Again the summations in the frequency interval $\varepsilon_n>0$ and $\varepsilon_n<0$ are $\omega_\nu$ independent and cancelled out by their mirror image diagrams,
thus we obtain
\begin{align}
	\Sigma^{\text{sj-DOS-(a+b)}}_{\bm{q}} = \frac{n_{\text{imp}}\alpha_{\text{so}}\nu V^2_0}{\varepsilon_F}\frac{\nu D}{8T} \Bigg\{\Theta(\Omega_k)\left[\psi'\left(\frac{1}{2}+\frac{2\omega_\nu+|\Omega_k|+Dq^2}{4\pi T}\right)-\psi'\left(\frac{1}{2}+\frac{|\Omega_k|+Dq^2}{4\pi T}\right)\right] \nonumber \\
	+ \Theta(-\Omega_k)\Theta(\Omega_k+\omega_{\nu-1})\left[\psi'\left(\frac{1}{2}+\frac{2\omega_\nu-|\Omega_k|+Dq^2}{4\pi T}\right)-\psi'\left(\frac{1}{2}+\frac{|\Omega_k|+Dq^2}{4\pi T}\right)\right]\Bigg\}.
\end{align}
Finally, we substitute $\Sigma^{\text{sj-DOS-(a+b)}}_{\bm{q}}(\Omega_k,\omega_\nu)$ into $Q^{\text{sj-DOS}}_{xy}(\omega_\nu)$, carry out the summation over $\Omega_k$, perform analytical continuation $\omega_\nu\rightarrow -i\omega$, and expand to linear order in $\omega$, to find 
\begin{equation}
\sigma^{\text{sj-DOS-(a+b)}}_{xy}=-\frac{14\zeta(3)}{\pi^2}\sigma_QD\varsigma_{\text{sj}}\int_{\bm{q}}\frac{1}{Dq^2+\tgl^{-1}},
\end{equation}
where we have introduced a dimensionless parameter $\varsigma_{\text{sj}}=n_{\text{imp}}\alpha_{\text{so}}\nu V^2_0/\varepsilon_F$ characterizing the rate of side-jump accumulation. 
Evaluations of the remaining diagrams in Fig.~\ref{fig:SJ-DOS} show that $\sigma^{\text{sj-DOS-c}}+\sigma^{\text{sj-DOS-d}}=\sigma^{\text{sj-DOS-a}}+\sigma^{\text{sj-DOS-b}}$, so that in total for the two-dimensional case 
\begin{equation}
	\sigma^{\text{sj-DOS}}_{xy}= 4\left(\sigma^{\text{sj-DOS-a}}+\sigma^{\text{sj-DOS-b}}+\sigma^{\text{sj-DOS-c}}+\sigma^{\text{sj-DOS-d}}\right)=-\frac{28\zeta(3)}{\pi^3}\sigma_Q\varsigma_{\text{sj}}\ln(T\tgl). 
\end{equation}
The relative importance of two extrinsic terms can be estimated as $\sigma^{\text{sk-DOS}}_{xy}/\sigma^{\text{sj-DOS}}_{xy}\sim(\varepsilon_F\tau) (\nu V_0)$.  For moderately strong impurity potential when, $\nu V_0\sim1$, skew scattering dominates in the metallic regime $\varepsilon_F\tau\gg1$. However, $\sigma^{\text{sk-MT}}_{xy}$ has an additional logarithmic in temperature enhancement as compared to $\sigma^{\text{sk-DOS}}_{xy}$. 


\section{AHE and ANE effects from interaction of superconducting fluctuations}\label{sec:nonlinear}

\begin{figure}
\centering
\includegraphics[width=0.55\linewidth]{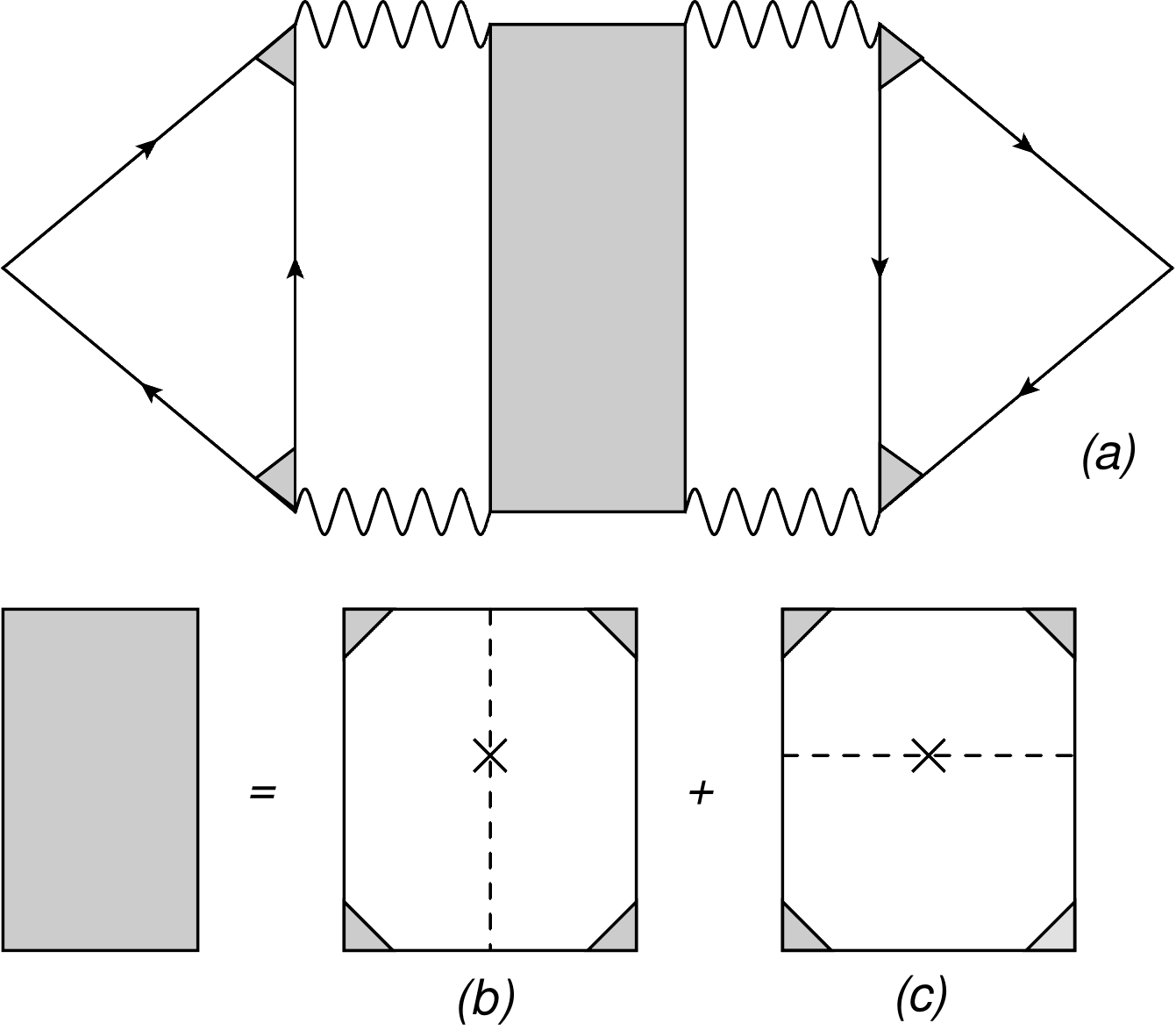}
\caption{Nonlinear Aslamazov-Larkin diagram-(a) that describes interaction of fluctuating superconducting modes. The quantum crossing of Cooperons is described by a Hikami box shown by a grey shaded rectangle whose analytical structure is defined by Eq. \eqref{eq:Hbox}. As further illustrated by panels (b)-(c) skew-scattering inside Hikami box leads to an asymmetry and thus renders finite AHE. The triangular blocks represent vector current vertices and defined by Eq. \eqref{eq:Be}. }
\label{fig:SK-AL}
\end{figure}

In order to appreciate the significant difference between Maki-Thompson ($\sigma^{\text{MT}}_{xy}$) and Aslamazov-Larkin ($\sigma^{\text{AL}}_{xy}$) terms in the context of anomalous Hall transport it will be useful to recall the difference between the two in the context of the usual Hall effect \cite{LV-Book}. The fluctuation processes of the MT and DOS types contribute to the renormalization of the diffusion coefficient so that they do not change Hall resistivity $\rho_{xy}=H/en$. For the Hall conductivity, however, one may write $\sigma_{xy}=\rho_{xy}\sigma^2_{xx}\approx \rho_{xy}\sigma^2_n+2\rho_{xy}\sigma_n\delta\sigma_{xx}$, with the correction $\delta\sigma_{xx}=\sigma^{\text{MT}}_{xx}+\sigma^{\text{DOS}}_{xx}$. Based on this simple argument it is expected that the relative fluctuation correction to Hall conductivity is twice  as large as the fluctuation correction to the diagonal component. Indeed, the microscopic diagrammatic analysis confirms this argument \cite{FET}, where one finds that $\sigma^{\text{MT}}_{xy}=2(\omega_c\tau)\sigma^{\text{MT}}_{xx}$, where $\omega_c$ is the cyclotron frequency. Curiously, we find essentially the same relationship between MT term in the skew scattering mechanism and diagonal MT contribution, at least in terms of their respective $T$-dependence, indeed from Eq. \eqref{eq:sigma-sk-MT} it follows that $\sigma^{\text{sk-MT}}_{xy}\simeq (\tau/\tau_{\text{sk}})\sigma^{\text{MT}}_{xx}$. The difference with the ordinary Hall effect is that the role of cyclotron frequency is replaced by the skew-scattering time $\omega_c\to\tau^{-1}_{\text{sk}}$, and we do not find the simple factor of two. The latter is clear, while having nonmagnetic disorder is sufficient to render finite MT correction to $\sigma_{xx}$, one needs to work beyond the leading Born order and use momentum dependent amplitude of scattering to generate finite corrections to $\sigma_{xy}$. In contrast, the AL process corresponds to an independent channel of the charge transfer that cannot be simply reduced to a renormalization of the diffusion coefficient. It contributes to the Hall conductivity only if particle-hole asymmetry of the Cooper pairs is present. For that reason $\sigma^{\text{AL}}_{xy}$ is not simply proportional to $\sigma^{\text{AL}}_{xx}$, like in the MT case, but rather one finds $\sigma^{\text{AL}}_{xy}=\frac{1}{3}(\omega_c\tau)\sigma^{\text{AL}}_{xx}\gamma_{\text{pha}}(T_c\tgl)$. The gauge invariance leads to the asymmetry factor in the form $\gamma_{\text{pha}}=\partial\ln T_c/\partial\ln \varepsilon_F$ \cite{AHL}. As a consequence, AL diagonal and Hall conductivities have distinct temperature dependences $\sigma^{\text{AL}}_{xx}\propto (T-T_c)^{-1}$ and $\sigma^{\text{AL}}_{xy}\propto (T-T_c)^{-2}$. At the level of the main AL diagram we do not find any possibilities to generate nonvanishing $\sigma^{\text{AL}}_{xy}$ by including skew scattering and side-jump scattering. Finite terms appear only at the next order that includes interaction of fluctuations. The leading process that we have identified is shown in Fig. \ref{fig:SK-AL}, which brings us to the realm of nonlinear superconducting fluctuations.              

In their early work Takayama-Maki \cite{Takayama-Maki} classified all the second-order corrections to the fluctuation-induced conductivity in an attempt to rigorously establish a control parameter of such a perturbative expansion. They identified five classes of diagrams that include: (I)-interaction corrections mixed between MT and DOS terms, (II)-renormalization of the electrical current vertex in the AL process, (III)-renormalization of the fluctuation propagator, (IV)-renormalization of electron propagator, and finally (V)-quantum interference of superconducting fluctuations of the AL type (see Fig. 3 in Ref. \cite{Takayama-Maki} for the diagrammatic representation of each class). A self-consistent calculation was further developed by Larkin-Ovchinnikov \cite{LO-NLAL}, who demonstrated that the Gaussian region of superconducting fluctuations is restricted to the temperature range $\sqrt{\text{Gi}}< (T-T_c)/T_c< 1$, while nonlinear effects dominate at $\text{Gi}< (T-T_c)/T_c<\sqrt{\text{Gi}}$ where perturbation theory is still well defined, see also Ref. \cite{AL-NLMTDOS}. Here $\text{Gi}\simeq 1/g_\Box\simeq1/(\varepsilon_F\tau)$ is the Ginzburg number and we used a standard definition in the transport context relating $\text{Gi}$ to a dimensionless conductance $g_\Box$ in 2D case (note that other definitions of $\text{Gi}$ exist in the context of thermodynamic properties, for example by relating fluctuation correction to specific heat and its jump at $T_c$, see Ref. \cite{LV-Book} for further details). These higher-order diagrams resurfaced numerous times in various other contexts, for example in the analysis of intrinsic dephasing in the Cooper channel \cite{Wolfle,Reizer-WL-MT}, in the study of current-voltage characteristics of superconducting tunnel and Josephson junctions \cite{Varlamov-Dorin,Reizer-Tunnel,AL-IcNoise-Tunnnel} and fluctuations in granular superconductors \cite{Lerner}, where tunneling plays a major role, mesoscopic fluctuations \cite{Zhou-Biagini,Skvortsov-GiantMeso}, and finally in calculations of the renormalization of the upper critical field \cite{Galitski-Hc2}. In this section we consider the most important diagrams from the class-(I) and (V) in the context of anomalous fluctuation-induced Hall and Nernst kinetic coefficients.           

\subsection{Skew-scattering Hikami quantum-crossing and Aslamazov-Larkin contribution to AHE}

The analysis of nonlinear AL term [Fig. \ref{fig:SK-AL}] requires consideration of two new elements in the diagrammatic technique. The first one is the triangular block of the electrical current vertex function, which we denote as $\bm{B}^{(e)}_{\bm{q}}$. The second one is a Hikami box \cite{Hikami} that captures quantum interferences processes in the crossings of cooperons, which we denote as $H_{\bm{qq}'}$. We begin with the former which is defined by the product of three Green's functions and two impurity vertex ladders summed up over the fermion frequency running in the loop and integrated over the electronic momentum. It reads explicitly     
\begin{align}\label{eq:Be}
\bm{B}^{(e)}_{\bm{q}}(\Omega_k,\omega_\nu)=2eT\sum_{\varepsilon_n}\lambda_{\bm{q}}(\varepsilon_n+\omega_\nu,\Omega_k-\varepsilon_n)\lambda_{\bm{q}}(\varepsilon_n,\Omega_k-\varepsilon_n) \nonumber \\ \times\int_{\bm{p}}\bm{v}_{\bm{p}}G_{\bm{p}}(\varepsilon_n+\omega_\nu)G_{\bm{p}}(\varepsilon_n)G_{\bm{q}-\bm{p}}(\Omega_k-\varepsilon_n).
\end{align}
Being a function of three frequencies and momentum, this vertex is fairly complicated, however in the classical region of fluctuations near $T_c$ its evaluation is relatively straightforward. The essential simplification comes from the separation of energy scales. Bosonic modes are pinned to the energy set by the pole structure of superconducting propagator $Dq^2\sim|\Omega_k|\sim T-T_c$. At the same time all fermionic modes are governed by the temperature $|\varepsilon_n|\sim T$. To capture the leading singularity near the transition, $T-T_c\ll T$, it is sufficient to evaluate the vertex function by setting all the bosonic frequencies to zero $\bm{B}^{(e)}_{\bm{q}}\to\bm{B}^{(e)}_{\bm{q}}(0,0)$. Recall that within the linear response Kubo analysis the external frequency is also set to zero in the end of the calculation. These approximations are not valid however in the regime of quantum fluctuations where more elaborate calculation of $\bm{B}^{(e)}_{\bm{q}}$ is required. As a result, near $T_c$ we can approximate
\begin{equation}
\bm{B}^{(e)}_{\bm{q}}=2eT\sum_{\varepsilon_n}
\lambda^2_{\bm{q}}(\varepsilon_n,-\varepsilon_n)
\int_{\bm{p}}\bm{v}_{\bm{p}}G^2_{\bm{p}}(\varepsilon_n)G_{\bm{q}-\bm{p}}(-\varepsilon_n).
\end{equation}
Transforming now momentum integration into the integral over fermionic dispersion, $\int_{\bm{p}}\ldots\to\nu\int d\xi_{\bm{p}}\langle\ldots\rangle$, where as in the previous cases averaging goes over the Fermi surface, and using Eq. \eqref{eq:G} we get
\begin{equation}
\bm{B}^{(e)}_{\bm{q}}=-2e\nu T\sum_{\varepsilon_n}
\lambda^2_{\bm{q}}(\varepsilon_n,-\varepsilon_n)\int^{+\infty}_{-\infty}
\frac{d\xi_{\bm{p}}}{(i\bar{\varepsilon}_n-\xi_{\bm{p}})^2}\left\langle\frac{\bm{v}}
{i\bar{\varepsilon}_n+\xi_{\bm{p}}-\bm{vq}}\right\rangle,
\end{equation}
where we expanded $\xi_{\bm{q}-\bm{p}}\approx\xi_{\bm{p}}-\bm{vq}$ in $G_{\bm{q}-\bm{p}}$. From here we need only the leading small-$q$ part of the vertex. Expanding the denominator to the linear order in $q$ and using Eq. \eqref{eq:lambda}, the above equation
transforms into
\begin{equation}
\bm{B}^{(e)}_{\bm{q}}=-2e\nu T\langle
\bm{v}(\bm{vq})\rangle\sum_{\varepsilon_n}\frac{|\bar{\varepsilon}_n|^2}{|\varepsilon_n|^2}
\int^{+\infty}_{-\infty}\frac{d\xi_{\bm{p}}}{(\xi^2_{\bm{p}}+\bar{\varepsilon}^2_n)^2}.
\end{equation}
The remaining $\xi_{\bm{p}}$-integration, followed by a $\varepsilon_n$-summation, can be completed in the closed form in terms of the digamma function
\begin{equation}\label{eq:Be-Ttau}
\bm{B}^{(e)}_{\bm{q}}=2e\bm{B}_{\bm{q}},\qquad \bm{B}_{\bm{q}}=2\nu D\tau
\bm{q}\left[\psi\left(\frac{1}{2}+\frac{1}{4\pi
T\tau}\right)-\psi\left(\frac{1}{2}\right)-\frac{1}{4\pi
T\tau}\psi'\left(\frac{1}{2}\right)\right].
\end{equation}
Focusing on the diffusive case only, $T\tau\ll1$, the above expression simplifies even further
\begin{equation}
\bm{B}_{\bm{q}}=-2\nu\eta \bm{q}\,,\qquad \eta=\pi D/8T\,.
\end{equation}

The Hikami box part of the AL diagram can be derived similarly by generalizing existing calculations known from the context of mesoscopic fluctuations to include skew-scattering amplitude \cite{Aronov-Hikami}. The resulting structure of the box is depicted in Fig. \ref{fig:SK-AL} (b-c) and its analytical structure has the form 
\begin{align}\label{eq:Hbox}
	& H^{\text{sk-(b)}}_{\bm{qq}'}=H^{\text{sk-(c)}}_{\bm{qq}'}=T\sum_{\varepsilon_n}\lambda^2_{\bm{q}}(\varepsilon_n,-\varepsilon_n)\lambda^2_{\bm{q}'}(\varepsilon_n,-\varepsilon_n) \nonumber\\
	&\times n_{\text{imp}}\int_{\bm{pp}'} V_{\bm{pp}'}V_{\bm{p}'\bm{p}}G^2_{\bm{p}}(\varepsilon_n)G_{\bm{q}-\bm{p}}(-\varepsilon_n)G^2_{\bm{p}'}(\varepsilon_n)G_{\bm{q}'-\bm{p}'}(-\varepsilon_n),
\end{align}
where the scattering amplitude, following Refs. \cite{Aronov-Hikami,AG-Knight}, is generalized as $V_{\bm{pp}'}=V_0\left[1-ib_{\bm{pp}'}(\hat{p}\times\hat{p}')_z\right]$ to allow for the arbitrary momentum dependence of the spin-orbit term to all orders in the scattering potential. Here $b_{\bm{pp}'}$ is in general a complex dimensionless function of momenta and we introduced a short-hand notation for a unit vector along momentum $\hat{p}=\bm{p}/p$. Full momentum dependence of the skew scattering cross section for strong potential can be calculated exactly in some model cases, for example in circular-barrier potential \cite{Raikh}. The product $V_{\bm{pp}'}V_{\bm{p}'\bm{p}}$ becomes, in the first order of $b_{\bm{pp}'}$, $V_{\bm{pp}'}V_{\bm{p}'\bm{p}}=2V^2_0\Im b_{\bm{pp}'}\,(\hat{p}\times\hat{p}')_z$, and for this reason calculation with Eq. \eqref{eq:V}, where constant amplitiude is assumed $b_{\bm{pp}'}\to\alpha_{\text{so}}$, is insufficient to capture skew-scattering in AL term. It can be immediately seen that one should expand Green's functions to the linear order order in $\bm{qq}'$ so that the angular average over the Fermi surface is nonvanishing. Specifically, we need the following expansion $G_{\bm{q}-\bm{p}}(-\varepsilon_n)\approx G_{-\bm{p}}(-\varepsilon_n)-(\bm{vq})G^2_{-\bm{p}}(-\varepsilon_n)$. The Hikami box then reduces to
\begin{align}
	& H^{\text{sk-(b)}}_{\bm{qq}'}=H^{\text{sk-(c)}}_{\bm{qq}'}=T\sum_{\varepsilon_n}\lambda^2_{\mathbf{q}\approx 0}(\varepsilon_n,-\varepsilon_n)\lambda^2_{\mathbf{q}'\approx 0}(\varepsilon_n,-\varepsilon_n) \Big\langle 2V^2_0\Im b_{\bm{pp}'}\,(\hat{p}\times\hat{p}')_z(\bm{v}_{\bm{p}}\bm{q})(\bm{v}_{\bm{p}'}\bm{q}') \Big\rangle_{\hat{p},\hat{p}'} \nonumber \\
	& \times n_{\text{imp}}\nu^2\int d\xi_{\bm{p}}d\xi_{\bm{p}'}G^2_{\bm{p}}(\varepsilon_n)G^2_{-\bm{p}}(-\varepsilon_n)G^2_{\bm{p}'}(\varepsilon_n)G^2_{-\bm{p}'}(-\varepsilon_n).   
\end{align}
Averaging over the directions of moment one has $\langle \cdots \rangle_{\hat{p},\hat{p}'}=\frac{2V^2_0v^2_F\Im b}{d^2}(\bm{q}\times\bm{q}')_z$, 
here $d$ is the dimensionality, and the notation for the momentum dependence in $b$ was suppressed for brevity. Next, integrating over dispersions $\xi_{\bm{p}}$ and $\xi_{\bm{p}'}$, one finds a factor of $(4\pi\nu\tau^3)^2$. Combining everything together we finally get for the Hikami box
\begin{align}\label{eq:H}
H^{\text{sk}}_{\bm{qq}'}=\frac{16}{3} \frac{n_{\text{imp}}\eta^2\nu^2 V^2_0 \Im b}{v^2_F T}(\bm{q}\times\bm{q}')_z=\frac{16}{3} \frac{\eta^2\nu}{v^2_F T\tau_{\text{sk}}}(\bm{q}\times\bm{q}')_z,
\end{align}
where the redefined skew-scattering time is $\tau^{-1}_{\text{sk}}=n_{\text{imp}}\nu V^2_0\Im b$, which applies only to AL diagram also in the context of the Nernst effect that will be discussed in the next section. As is well known, vanishing of the Hikami box in the $q\to0$ limit corresponds to the normalization of the probability of quantum diffusion that essentially follows from the conservation of the particle number. 
 
With these ingredients at hand we are prepared now to consider the electromagnetic response function of the AL term. It reads 
\begin{equation}
Q^{\text{sk-AL}}_{xy}=T^2\int_{\bm{qq}'}\big[\bm{B}^{(e)}_{\bm{q}}\big]_x\big[\bm{B}^{(e)}_{\bm{q}'}\big]_yH^{\text{sk}}_{\bm{qq}'}
\sum_{\Omega_k}L_{\bm{q}}(\Omega_k)L_{\bm{q}}(\Omega_k+\omega_\nu)
\sum_{\Omega'_k}L_{\bm{q}'}(\Omega'_k)L_{\bm{q}'}(\Omega'_k+\omega_\nu)
\end{equation}
The frequency summation over $\Omega_k$ can be evaluated via the contour integral in the complex plane with the $\coth$-function that has simple poles and unity residues as follows    
\begin{align}
	T\sum_{\Omega_k}L_{\bm{q}}(\Omega_k)L_{\bm{q}}(\Omega_k+\omega_\nu)=\oint \frac{d\Omega}{4\pi i}\coth(\Omega/2T)L_{\bm{q}}(-i\Omega)L_{\bm{q}}(-i\Omega+\omega_\nu)
	\nonumber \\
	=\int^{+\infty}_{-\infty}\frac{d\Omega}{2\pi}\coth (\Omega/2T)\left[L^R_{\bm{q}}(-i\Omega+\omega_\nu)-L^A_{\bm{q}}(-i\Omega-\omega_\nu)\right] \Im L^R_{\bm{q}}(-i\Omega),
\end{align}
where one has to account for the breaks of analyticity of $L_{\bm{q}}(\Omega_k)$ as a function of frequency when transforming the integral. Thus one gets a particular combinations of retarded and advanced components. The summation over $\Omega'_k$ is completely analogous. As the next step we analytically continue the resulting expression $\omega_\nu\to-i\omega$, and expand $Q^{\text{sk-AL}}_{xy}$ to the linear order in external frequency. In performing frequency $(\Omega,\Omega')$-integrals we notice that the most relevant region is $\Omega,\Omega'\sim T-T_c$ so that one can safely expand $\coth(\Omega/2T)\approx 2T/\Omega$ as singular parts are cancelled by extra powers of $\Omega$ in $\Im L^{R/A}_{\bm{q}}$.  
In other words we can approximate 
\begin{equation}\label{eq:coth-L-approx}
\coth(\Omega/2T)\Im L^R_{\bm{q}}(-i\Omega)\approx-\frac{16T^2}{\pi\nu}\frac{1}{(Dq^2+\tgl^{-1})^2+\Omega^2}.
\end{equation}
Then at the intermediate step we find
\begin{equation}
\sigma^{\text{sk-AL}}_{xy}=\frac{32\pi e^2\eta^4}{3\nu v^2_F\tau_{\text{sk}}}\int_{\bm{qq}'}\frac{q_xq'_y(\bm{q}\times\bm{q}')_z}{\big[\xi^2 q^2+t\big]^3\big[\xi^2 q'^2+t\big]^2},\quad t=\frac{T-T_c}{T_c}.
\end{equation} 
Here we introduced coherence length $\xi=\sqrt{\pi D/8T_c}$ in the diffusive limit. In the remaining momentum integrations one notices that $q'$-integral is logarithmically divergent in the upper limit. However, this divergence is easily cured by retaining $q'$ dependence of the corner cooperons in the Hikami box, originating from the zeroth Matsubara frequency term of $\lambda_{\bm{q}'}(\varepsilon_n,-\varepsilon_n)$, which ads an additional Lorentzian factor $[\xi^2 q'^2+\pi^2/2]^{-2}$ inside the integrand of the last formula. As a result, with the logarithmic accuracy we find  
\begin{equation}\label{eq:sigma-sk-AL}
\sigma^{\text{sk-AL}}_{xy}=\frac{\sigma_Q}{12\varepsilon_F\tau} (\tau/\tau_{\text{sk}})(T_c\tgl)\ln(T_c\tgl). 
\end{equation}
This result should be compared to $\sigma^{\text{sk-MT}}_{xy}$ from Eq. \eqref{eq:sigma-sk-MT}. Coincidentally both terms have the same temperature dependence, however, nonlinear AL contribution is parametrically smaller in $1/(\varepsilon_F\tau)\ll1$. We will show in the section below that nevertheless quantum interference of superconducting fluctuations captured by nonlinear AL process plays an important role in the thermo-magnetic response where MT term vanishes.  

\begin{figure}
\centering
\includegraphics[width=0.55\linewidth]{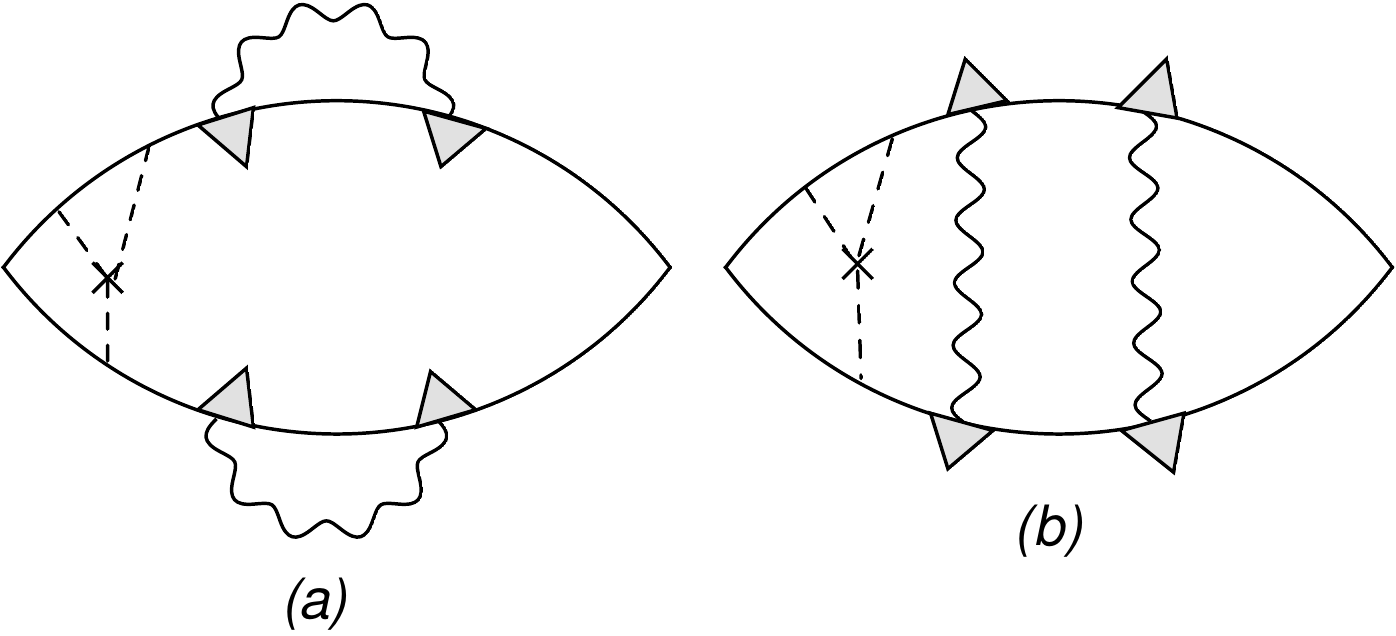}
\caption{The second-loop diagrams for DOS-(a) and MT-(b) corrections to the anomalous Hall conductivity in the skew
scattering mechanism near $T_c$ due to interaction of superconducting fluctuations. These diagrams show a basic skeleton and other similar diagrams can be generated by 
reordering of impurity lines in the skew-scattering amplitude. } 
\label{fig:SK-MTDOS}
\end{figure}

\subsection{Nonlinear Maki-Thompson and density of states interference corrections to AHE}

Interaction of fluctuations occurs also via the second order DOS and MT interference processes. Two representative diagrams are depicted in Fig. \ref{fig:SK-MTDOS}. We concentrate on the skew-scattering mechanism only as side-jumps, although possible, are expected to be smaller for realistic impurities. We already elaborated on this point in Sec. \ref{sec:sj}. For the nonlinear density of states effect [Fig. \ref{fig:SK-MTDOS}(a)] the electromagnetic response correlation function in the Kubo formula is given by  
\begin{equation}
Q^{\text{sk-DOS-nl}}_{xy}(\omega_\nu)=2e^2T^2\int_{\bm{qq}'}\sum_{\Omega_k\Omega'_l} L_{\bm{q}}(\Omega_k)L_{\bm{q}'}(\Omega'_l)\Sigma^{\text{sk-DOS-nl}}_{\bm{q}\bm{q}'}(\Omega_k,\Omega'_l,\omega_\nu),
\end{equation}
The corresponding self-energy block is 
\begin{equation}
 \Sigma^{\text{sk-DOS-nl}}_{\bm{qq}'}(\Omega_k,\Omega'_l,\omega_\nu) = T\sum_{\varepsilon_n}\lambda^2_{\bm{q}}(\varepsilon_n,\Omega_k-\varepsilon_n)\lambda^2_{\bm{q}'}(\varepsilon_{n+\nu},\Omega_k-\varepsilon_{n+\nu})J^{\text{sk-DOS-nl}}_{xy}
\end{equation}
where 
\begin{align}
J^{\text{sk-DOS-nl}}_{xy}=
n_{\text{imp}}\nu^3 \langle v_{\bm{p},x}v_{\bm{k},y}V_{\bm{p}\bm{k}'}V_{\bm{k}'\bm{k}}V_{\bm{k}\bm{p}}\rangle 
\int d\xi_{\bm{p}} G_{\bm{p}}(\varepsilon_n)G_{\bm{p}}(\varepsilon_{n+\nu}) \int d\xi_{\bm{k}'} G_{\bm{k}'}(\varepsilon_{n+\nu})\nonumber \\ 
\times \int d\xi_{\bm{k}} G^2_{\bm{k}}(\varepsilon_n)G_{-\bm{k}}(\Omega_k-\varepsilon_n)G^2_{\bm{k}}(\varepsilon_{n+\nu})G_{-\bm{k}}(\Omega'_l-\varepsilon_{n+\nu}).
\end{align}
After all the standard technical steps of carrying out integrals over the fermionic dispersions, performing angular averages, completing summations over frequencies followed by analytical continuation we obtain for the conductivity 
\begin{align}
	\sigma^{\text{sk-DOS-nl}}_{xy}&=24\pi^3e^2\nu D(\tau/\tau_{\text{sk}})\int_{\bm{qq}'}\int \frac{d\varepsilon}{2\pi} \frac{\partial}{\partial \varepsilon} \left(\tanh{\frac{\varepsilon}{2T}}\right) \nonumber \\ &\times\int \frac{d\Omega d\Omega'}{(2\pi)^2}  \frac{\Im L^R_{\bm{q}}(-i\Omega)\coth(\Omega/2T)}{\big[Dq^2+i(2\varepsilon-\Omega)\big]^2}\frac{\Im L^R_{\bm{q}'}(-i\Omega')\coth(\Omega'/2T)}{\big[Dq'^2-i(2\varepsilon-\Omega')\big]^2}. 
\end{align}
It is worthwhile noting that even though we labeled this contribution as nonlinear DOS effect its analytical structure with the mixed product of advanced and retarded cooperons in fact corresponds to the anomalous MT term. We thus cautiously state that already at the second order there is no simple classification of various terms just as DOS and MT so that the terminology here is somewhat symbolic. In order to extract the most singular part of the above expression it is sufficient to approximate $\partial_\varepsilon\tanh(\varepsilon/2T)\approx1/2T$ under the integral and use Eq. \eqref{eq:coth-L-approx}. Then all remaining frequency and momentum integrations become elementary and can be done in the closed form by residues. We find the result 
\begin{equation}\label{eq:sigma-sk-DOS-nl}
\sigma^{\text{sk-DOS-nl}}_{xy}\simeq\frac{\sigma_Q}{\varepsilon_F\tau} (\tau/\tau_{\text{sk}})(T_c\tgl)^3
\end{equation} 
where we omitted overall numerical pre-factor which is of the order of unity. The self-energy part of the nonlinear MT diagram-(b) in Fig. \ref{fig:SK-MTDOS} is of the form 
\begin{align}
\Sigma^{\text{sk-MT-nl}}_{\bm{qq}'}(\Omega_k,\Omega'_l,\omega_\nu) = T\sum_{\varepsilon_n}\lambda_{\bm{q}}(\varepsilon_n,\Omega_k-\varepsilon_n)\lambda_{\bm{q}}(\varepsilon_{n+\nu},\Omega_k-\varepsilon_{n+\nu})\nonumber \\ 
\times\lambda_{\bm{q}'}(-\varepsilon_n,\Omega'_l+\varepsilon_n)\lambda_{\bm{q}'}(-\varepsilon_{n+\nu},\Omega'_l+\varepsilon_{n+\nu})J^{\text{sk-MT-nl}}_{xy}
\end{align}
where 
\begin{align}
J^{\text{sk-MT-nl}}_{xy}=n_{\text{imp}}\nu^3\langle v_{\bm{p},x}v_{\bm{k},y}V_{\bm{p}\bm{k}'}V_{\bm{k}'\bm{k}}V_{\bm{k}\bm{p}}\rangle\int d\xi_{\bm{p}} d\xi_{\bm{k}'} G_{\bm{p}}(\varepsilon_n)G_{\bm{p}}(\varepsilon_{n+\nu}) G_{\bm{k}'}(\varepsilon_{n+\nu})\nonumber \\
 \times \int d\xi_{\bm{k}} G^2_{\bm{k}}(\varepsilon_n)G_{-\bm{k}}(\Omega_k-\varepsilon_n)G^2_{\bm{k}}(\varepsilon_{n+\nu})G_{-\bm{k}}(\Omega'_l-\varepsilon_{n+\nu}). 
\end{align}
This translates to the conductivity correction in the form 
\begin{align}
	\sigma^{\text{sk-MT-nl}}_{xy}&=16\pi^3e^2\nu D (\tau/\tau_{\text{sk}})\int_{\bm{qq}'}\int \frac{d\varepsilon}{2\pi} \frac{\partial}{\partial \varepsilon} \left(\tanh{\frac{\varepsilon}{2T}}\right) \nonumber \\ &\times\int \frac{d\Omega d\Omega'}{(2\pi)^2}  \frac{\Im L^R_{\bm{q}}(-i\Omega)\coth(\Omega/2T)}{\big[(Dq^2+\tau^{-1}_\phi)^2+(2\varepsilon-\Omega)^2\big]}\frac{\Im L^R_{\bm{q}'}(-i\Omega')\coth(\Omega'/2T)}{\big[(Dq'^2+\tau^{-1}_\phi)^2+(2\varepsilon-\Omega')^2\big]}
\end{align}
where we retained the dephasing time in the cooperons as it plays an important role in the regularization of this contribution. The respective shift of $T_c$ is implicit in the pair propagators. 
In the limit when depairing is weak, $\tau_\phi\gg\tgl$, and with the logarithmic accuracy we extract the temperature dependence in the form up to a numerical factor   
\begin{equation}\label{eq:sigma-sk-MT-nl}
\sigma^{\text{sk-MT-nl}}_{xy}\simeq\frac{\sigma_Q}{\varepsilon_F\tau}(\tau/\tau_{\text{sk}})(T_c\tgl)^3\ln^2(\tau_\phi/\tgl).
\end{equation} 
We conclude that nonlinear terms of MT and DOS type are more singular than AL correction Eq. \eqref{eq:sigma-sk-AL}. Furthermore, comparing Eqs. \eqref{eq:sigma-sk-DOS-nl} and \eqref{eq:sigma-sk-MT-nl} with Eq. \eqref{eq:sigma-sk-MT} we see that nonlinear terms become more important than linear corrections at temperatures $(T-T_c)/T_c<1/\sqrt{\varepsilon_F\tau}$. Note however that at this scale $\sigma^{\text{sk-MT-nl}}_{xy}$ is still smaller than its normal state value $\sigma^{\text{n}}_{xy}\sim(\tau/\tau_{\text{sk}})\sigma_n$ in a parameter $\sigma^{\text{sk-MT-nl}}_{xy}/\sigma^{\text{n}}_{xy}\sim1/\sqrt{\varepsilon_F\tau}\ll1$ so that perturbative expansion in fluctuations is still justified.  This confirms an earlier assertion that nonlinear fluctuations could play a dominant role in determining transport characteristics sufficiently close to $T_c$ at $\text{Gi}< (T-T_c)/T_c<\sqrt{\text{Gi}}$.  

\subsection{Nonlinear Aslamazov-Larkin mechanism of ANE}

Thermal fluctuations of the superconducting order parameter also contribute to thermoelectric responses. This includes both longitudinal thermopower, Seebeck and Peltier effects, and also transversal responses such as Nernst and Ettingshausen effects. These phenomena can be viewed through the prism of Onsager reciprocal relations for kinetic coefficients as superconducting fluctuations transport heat current in response to an electric field, or conversely transport electric current in response to the temperature gradient applied to the system. The transversal responses obviously require external magnetic field and in this section we will look at their anomalous counterparts.  
  
Thermoelectric coefficients  can be also computed from the Kubo formula where we need to know the mixed thermal current-electrical current response function, which we label as $K_{ij}(\omega_\nu)$. Provided that the analytically continued retarded function $K^R_{ij}(\omega)$ is known the tensor of thermoelectrical conductivities follows $\beta_{ij}=-\frac{1}{T}\lim_{\omega\to0}\Im K^R_{ij}/\omega$.  We briefly recapitulate computation of the diagonal component of $\beta_{ij}$ first from the AL diagram \cite{LV-Book} and then generalize it to the case of off-diagonal component. The thermal current block of the AL diagram corresponds to the following triangular vertex    
 \begin{align}\label{eq:Bh}
\bm{B}^{(h)}_{\bm{q}}(\Omega_k,\omega_\nu)=T\sum_{\varepsilon_n}\frac{i(\varepsilon_{n+\nu}+\varepsilon_n)}{2}\lambda_{\bm{q}}(\varepsilon_n+\omega_\nu,\Omega_k-\varepsilon_n)\lambda_{\bm{q}}(\varepsilon_n,\Omega_k-\varepsilon_n) \nonumber \\ \times\int_{\bm{p}}\bm{v}_{\bm{p}}G_{\bm{p}}(\varepsilon_n+\omega_\nu)G_{\bm{p}}(\varepsilon_n)G_{\bm{q}-\bm{p}}(\Omega_k-\varepsilon_n).
\end{align}
Ussishkin \cite{Ussishkin} analyzed this function and showed its connection to the corresponding electrical block which was introduced earlier in Eq. \eqref{eq:Be}. Specifically one finds 
$\bm{B}^{(h)}_{\bm{q}}=[i(\Omega_k+\omega_\nu/2)/2e]\bm{B}^{(e)}_{\bm{q}}$ which yields the thermoelectric response kernel 
\begin{equation}\label{eq:K-AL}
K^{\text{AL}}_{xx}=T\int_{\bm{q}}\sum_{\Omega_k}\big[\bm{B}^{(h)}_{\bm{q}}\big]_x\big[\bm{B}^{(e)}_{\bm{q}}\big]_xL_{\bm{q}}(\Omega_k)L_{\bm{q}}(\Omega_k+\omega_\nu).
\end{equation} 
Summation over the Matsubara frequency $\Omega_k$ and analytical continuation follows the same way as in the case of the conductivity calculation, so that we obtain
\begin{equation}
\beta^{\text{AL}}_{xx}=\frac{e}{2\pi T^2}\int_{\bm{q}}\big[\bm{B}_{\bm{q}}\big]^2_x\int\frac{\Omega d\Omega}{\sinh^{2}(\Omega/2T)}\big[\Im L^R_{\bm{q}}(-i\Omega)\big]^2.
\end{equation}  
It is seen here that for the pair propagator in the form of Eq. \eqref{eq:L} frequency integration in the above formula gives identical zero for $\beta_{xx}$ since $[\Im L^R_{\bm{q}}]^2$ is an even function of $\Omega$ whereas the rest of the integrand is odd. Thus one needs a more general expression that accounts for the particle-hole asymmetry \cite{AHL}   
\begin{equation}
L_{\bm{q}}(\Omega_k)=-\frac{1}{\nu}\frac{1}{\pi Dq^2/8T+t+\pi|\Omega_k|/8T+\Upsilon_\Omega},\quad \Upsilon_\Omega=(i\Omega_k/2T_c)(\partial T_c/\partial\varepsilon_F).
\end{equation}  
The asymmetry factor $\Upsilon_\Omega$ in particular accounts for the variation of the density of states at the Fermi surface. When expanding $L^R_{\bm{q}}$
 to the linear in $\Upsilon_\Omega$ order produces
 \begin{equation}
\beta^{\text{AL}}_{xx}=-\frac{e\nu}{\pi T^2}\int_{\bm{q}}\big[\bm{B}_{\bm{q}}\big]^2_x\int\frac{\Omega \Upsilon_\Omega d\Omega}{\sinh^{2}(\Omega/2T)}
\big[\Im L^R_{\bm{q}}(-i\Omega)\big]\Im\big[L^R_{\bm{q}}(-i\Omega)\big]^2
\end{equation}  
where now both propagators have been taken at $\Upsilon_\Omega\to0$. In the end, we recover a well known result \cite{LV-Book} 
\begin{equation}
\beta^{\text{AL}}_{xx}\simeq\beta_Q\varsigma_{\text{pha}}\ln(T_c\tgl),\quad \varsigma_{\text{pha}}=\frac{\partial T_c}{\partial\varepsilon_F} 
\end{equation}
where $\beta_Q=e/(2\pi\hbar)$ is the quantum unit of the thermoelectric coefficient.

One can generalize now Eq. \eqref{eq:K-AL} to include the Hikami box with skew scattering Eq. \eqref{eq:H} that induces the transversal part of the thermoelectric response. This is described by the following expression 
 \begin{equation}
K^{\text{sk-AL}}_{xy}=T^2\int_{\bm{qq}'}\big[\bm{B}^{(h)}_{\bm{q}}\big]_x\big[\bm{B}^{(e)}_{\bm{q}'}\big]_yH^{\text{sk}}_{\bm{qq}'}
\sum_{\Omega_k}L_{\bm{q}}(\Omega_k)L_{\bm{q}}(\Omega_k+\omega_\nu)
\sum_{\Omega'_k}L_{\bm{q}'}(\Omega'_k)L_{\bm{q}'}(\Omega'_k+\omega_\nu)
\end{equation}
By repeating the steps that us lead to the equation for $\sigma^{\text{AL}}_{xx}$ we find instead 
\begin{equation}
\beta^{\text{sk-AL}}_{xy}=
\frac{2^8e\varsigma_{\text{pha}}\eta^4}{3\pi\nu v^2_F\tau_{\text{sk}}} \int_{\bm{qq}'} \frac{q_xq'_y(\mathbf{q}\times\mathbf{q}')_z}{\big[\xi^2 q^2+t\big]^2\big[\xi^2 q'^2+t\big]^2}.
 \end{equation}
and finally for the temperature dependence of the anomalous Nernst coefficient   
\begin{equation}\label{eq:beta-sk-AL}
\beta^{\text{sk-AL}}_{xy}\simeq\frac{\beta_Q}{\varepsilon_F\tau}\varsigma_{\text{pha}}(\tau/\tau_{\text{sk}})\ln^2(T_c\tgl). 
\end{equation}  

There are several interesting aspects of the results obtained in this section that deserve further discussion. Apparently there exists a substantial difference in how particle-hole asymmetry affects fluctuation-induced transversal response in superconductors. The ordinary Hall conductivity requires finite $\varsigma_{\text{pha}}$ \cite{AHL} whereas Nernst effect is not \cite{Ussishkin}. It is the other way around for the anomalous coefficients as both skew-scattering and asymmetry are required for finite thermoelectric response Eq. \eqref{eq:beta-sk-AL}, whereas anomalous Hall conductivity is already presents at vanishing $\varsigma_{\text{pha}}$ Eq. \eqref{eq:sigma-sk-AL}. In addition, the sensitivity of transversal transport coefficients to particle-hole asymmetry makes them useful for diagnostics of effects related to Fermi surface reconstruction. Of particular interest are diagnostics for electronic Lifshitz-type transition associated with a change of topology of the Fermi surface \cite{Varlamov-ETT} that can be induced by doping, changing the impurity concentration, applying pressure or stress.        


\section{Fluctuation-induced transport in superconductors at strong-coupling}\label{sec:eliashberg}

The substance of this section is devoted to an outline of main ideas and discussion of technical details needed to extend existing theory of superconducting fluctuations to the regime of strong coupling in Eliashberg formulation \cite{Eliashberg}. In the context of the present study, the motivation for the need of this elaborate framework is clear. As we summarized in the opening section, most of the transport anomalies of interest were observed in unconventional superconductors where BCS weak coupling description is not expected to hold. However, since microscopic origin of electronic pairing glue is still been actively debated practically for all the existing unconventional systems, it makes sense to begin with the canonical electron-phonon mechanism. In principle, it should be then possible to extend the method further for the case of pairing mediated by a composite low-energy boson.       

Eliashberg's approach has been applied by Bulaevskii-Dolgov \cite{Bulaevskii-Dolgov} and Narozhny \cite{Narozhny} to calculate the fluctuation effects in the specific heat $\delta c$ and diamagnetic susceptibility $\delta\chi$ above $T_c$. They concluded that the resulting temperature dependence of both $\delta c$ and $\delta\chi$ is the same as in the BCS weak coupling limit \cite{LV-Book}, namely $(\delta c, \delta\chi)\propto 1/t$ in 2D case, and $\propto 1/\sqrt{t}$ in 3D, but in the extreme limit of the theory fluctuations are enhanced by a large dimensionless parameter $\lambda\gg1$. This implies that at strong coupling fluctuational region widens which is at least qualitatively consistent with numerous observations where fluctuations survive relatively far away from $T_c$. The dimensionless coupling constant $\lambda$ can be re-expressed as the ratio of some effective interaction $g$ in the model and the Debye frequency $\lambda=(g/\omega_D)^2$, and related to the integrated bosonic spectral function. For large $\lambda$, as demonstrated by Allen and Dynes \cite{Allen-Dynes}, the critical temperature scales within a numerical factor as $T_c\simeq \omega_D\sqrt{\lambda}\sim g$. At zero temperature the energy gap scales with $T_c$, $\Delta\sim T_c$ similar to the BCS limit albeit the proportionality coefficient is different. 

The method of Ref. \cite{Bulaevskii-Dolgov}, where an effective Ginzburg-Landau description in strong coupling was derived, can not be directly applied to the calculation fluctuation-induced corrections in transport coefficients as one actually needs Keldysh type representation in terms of real energies \cite{AL-TDGL}. This representation is needed for both vertex functions of electromagnetic response kernels and pair propagator itself. Fortunately a progress has been made in this direction as analytical continuation of Eliashberg gap equations has been implemented in the works of Marsiglio-Schossmann-Carbotte \cite{MSC} and Karakozov-Maksimov-Mikhailovskii \cite{KMM}, see also a detailed study of Combescot \cite{Combescot}. To make use of these advances, in principle, one can proceed in two complimentary ways. The first route is to work in the path integral representation to construct Luttinger-Ward functional. The saddle point of the corresponding action will be a linearized Eliashberg gap equation. The expansion around this solution to the quadratic order in low-energy modes of pair excitations should lead to the propagator of superconducting fluctuations. We choose to follow an alternative way, which is more naturally connected to the diagrammatic approach employed in this paper, and consider Bethe-Salpeter equation for the irreducible part of the two-particle Green's function $\Gamma^C_{\bm{pp'q}}$ \cite{LV-Book,Bulaevskii-Dolgov,Narozhny}. To avoid further complications with the disorder averaging we concentrate on the clean case and focus only on the strong coupling effects in fluctuations. We assume that electron-phonon scattering relaxes electronic momentum and thus gives finite normal state conductivity. Also having in mind applications to superconducting films, we further assume that film thickness is such that spectra of electrons and phonons are 3D while the SC fluctuations are effectively 2D close to $T_c$.   

The irreducible two-particle vertex obeys the integral equation 
\begin{equation}\label{eq:Gamma-Eliashberg}
\Gamma^C_{\bm{pp'q}}(\varepsilon_n,\varepsilon_m,\Omega_k)=D_{\bm{p}-\bm{p}'}(\varepsilon_n-\varepsilon_m)+T\sum_{\varepsilon_l}\int_{\bm{k}}D_{\bm{p-k}}(\varepsilon_l-\varepsilon_n)
G_{\bm{q}-\bm{k}}(\varepsilon_l+\Omega_k)G_{\bm{k}}(-\varepsilon_l)\Gamma^C_{\bm{kp'q}}(\varepsilon_l,\varepsilon_m,\Omega_k)
\end{equation}
where $D_{\bm{q}}(\omega_k)$ is a propagator for a phonon. In the Eliashberg theory, electron Green's function $G^{-1}_{\bm{p}}(\varepsilon_n)=i\varepsilon_n-\xi_{\bm{p}}+\Sigma(\varepsilon_n)$ self-consistently contains self-energy mediated by retardation of interactions with phonons which is also an integral equation of Dyson type 
\begin{equation}\label{eq:Sigma}
\Sigma(\varepsilon_n)=T\sum_{\omega_k}\int_{\bm{q}}D_{\bm{q}}(\omega_k)G_{\bm{p}-\bm{q}}(\varepsilon_n-\omega_k).
\end{equation}  
It can be directly seen that by replacing phonon propagator with a constant in Eq. \eqref{eq:Gamma-Eliashberg}, in a spirit of the weak-coupling approach with point interactions, one immediately recovers Eq. \eqref{eq:L-Dyson} for the pair-propagator. Note that in the absence of impurity scattering the impurity ladder function $\lambda_{\bm{q}}(\varepsilon,\varepsilon')$ in Eq. \eqref{eq:L-Dyson} should be replaced by unity.  In general, solution to coupled integral equations for $\Gamma^C$ and $\Sigma$ poses a daunting problem. Near $T_c$, however, a major simplification is possible that makes this problem tractable analytically. Indeed, at $T=T_c$ as $q,\Omega\to0$ vertex function $\Gamma^C$ has a pole. At temperature close to $T_c$, the proximity to $T-T_c$ at finite momentum $q$ and energy $\Omega_k$ sets the typical scale for bosonic fluctuating modes. At the same time, fermionic modes are fast in this regime with typical excitations of the order of temperature $\varepsilon\sim T$. This separation of energy scales, $\varepsilon\gg \Omega$, makes it possible to postulate a physically motivated ansatz   
\begin{equation}\label{eq:Gamma-approx}
\Gamma^C_{\bm{pp'q}}(\varepsilon_n,\varepsilon_m,\Omega_k)=\Delta(\varepsilon_n)\Delta(\varepsilon_m)L_{\bm{q}}(\Omega_k)
\end{equation}
where one singles out singular part of the vertex into a pair-propagator that is superimposed with the product of smooth amplitude eigenfunctions $\Delta(\varepsilon_n)$. As we will see momentarily the latter can be adjusted to fulfill exactly the linearized homogeneous Eliashberg equation for $T=T_c$. In this approximation electronic momenta $\bm{p}, \bm{p}'$ are set to Fermi momentum and Eliashberg interaction function is defined in a standard way by averaging phonon propagator over the Fermi surface 
\begin{equation}\label{eq:Lambda}
\Lambda(\varepsilon_n)=\int^{\infty}_{0}\frac{2\omega d\omega}{\varepsilon^2_n+\omega^2}\alpha^2(\omega)F(\omega)
\end{equation}
where $\alpha^2(\omega)F(\omega)$ is the phonon spectral function introduced by Eliashberg. As demonstrated by  Combescot, the actual shape of $\alpha^2(\omega)F(\omega)$ is irrelevant at strong coupling, as the bosonic spectral function enters into the theory only via the effective dimensionless interaction parameter $\lambda=\frac{1}{\langle \omega^2\rangle}\int 2\omega\alpha^2(\omega)F(\omega)d\omega$ where parametrically $\langle \omega^2\rangle\sim \omega^2_D$. With the trial form of $\Gamma^C$ from Eq. \eqref{eq:Gamma-approx} the Bethe-Salpeter equation \eqref{eq:Gamma-Eliashberg} reduces to the Fredholm integral equation of the second kind with the separable kernel. This equation can be solved exactly. Indeed, we bring Eq. \eqref{eq:Gamma-approx} into Eq. \eqref{eq:Gamma-Eliashberg}, multiply both sides by the product $\Delta(\varepsilon_n)\Delta(\varepsilon_m)$ and sum over both fermionic Matsubara frequencies. For the amplitude functions with the Plancherel normalization to unity this gives an algebraic equation for the pair propagator that is solved by 
\begin{equation}
L_{\bm{q}}(\Omega_k)=\frac{N}{1-P_{\bm{q}}(\Omega_k)}.
\end{equation}    
Here norm in the numerator is given by 
\begin{equation}\label{eq:N}
N=\sum_{\varepsilon_n\varepsilon_m}\Delta(\varepsilon_n)\Delta(\varepsilon_m)D(\varepsilon_n-\varepsilon_m),
\end{equation}
whereas polarization operator in the denominator is determined by 
\begin{equation}\label{eq:P}
P_{\bm{q}}(\Omega_k)=T\sum_{\varepsilon_n\varepsilon_l}\int_{\bm{k}}\Delta(\varepsilon_l)\Delta(\varepsilon_n)D(\varepsilon_l-\varepsilon_n)G_{\bm{k}}(-\varepsilon_l)G_{\bm{q}-\bm{k}}(\varepsilon_l+\Omega_k).
\end{equation}
It readily follows that $P_{\bm{q}\to0}(\Omega_k\to0)=1$ at $T=T_c$ provided that $\Delta(\varepsilon_n)$ satisfies an equation 
\begin{equation}\label{eq:Delta}
\Delta(\varepsilon_n)=T_c\sum_{\varepsilon_n}\int_{\bm{k}}
\Delta(\varepsilon_l)D(\varepsilon_l-\varepsilon_n)G_{\bm{k}}(-\varepsilon_l)G_{-\bm{k}}(\varepsilon_l)=-\frac{iT_c}{2\pi}
\sum_{\varepsilon_l}\Lambda(\varepsilon_l-\varepsilon_n)\frac{\Delta(\varepsilon_l)}{i\varepsilon_l+\Sigma(\varepsilon_l)}
\end{equation} 
which coincides with the linearized Eliashberg equation for $T_c$. In this model self-energy can be also expressed in therms of $\Lambda$-function. From Eq. \eqref{eq:Sigma} it follows that $\Sigma(\varepsilon_n)=-\frac{iT_c}{2\pi}\big[\Lambda(0)+2\sum^{n}_{l=1}\Lambda(\varepsilon_l)\big]$. For the purpose of conductivity calculation we need a current vertex function. For the AL diagram the corresponding triangular block is given by 
\begin{equation}\label{eq:Be-Eliashberg}
\bm{B}^{(e)}_{\bm{q}}(\Omega_k,\omega_\nu)=2eT\sum_{\varepsilon_n}\bm{v}_{\bm{p}}G_{\bm{p}}(\varepsilon_n+\omega_\nu)G_{\bm{p}}(\varepsilon_n)G_{-\bm{p}+\bm{q}}(\Omega_k-\varepsilon_n)\Delta(\varepsilon_n+\omega_\nu)\Delta(\varepsilon_n).
\end{equation} 
This expression should be compared to the Eq. \eqref{eq:Be} in the weak coupling formulation. The difference is twofold, in the clean case $\lambda_{\bm{q}}\to1$ in Eq. \eqref{eq:Be}, however instead the strong coupling effects add two factors of dynamical amplitude function $\Delta(\varepsilon_n)$ due to parametrization of the irreducible vertex $\Gamma^C$ in Eq. \eqref{eq:Gamma-approx}. 

As the next step we need to analytically continue Eqs. \eqref{eq:Sigma}, \eqref{eq:N}--\eqref{eq:Delta} from the Matsubara to real frequencies. Following the prescription of Ref. \cite{KMM} and expanding Eq. \eqref{eq:P} at low frequencies $\Omega_k$ and momenta $q$ we establish a functional form of the pair propagator
\begin{equation}
P^R_{\bm{q}}(-i\Omega)=1+p_qq^2+p_t t-ip_\omega\Omega,\quad t=(T-T_c)/T_c,
\end{equation}   
where a set of functions $p_{q,t,\omega}$ has an extremely complicated form that can be asymptotically estimated for $\lambda\gg1$. With this form of $P^R_{\bm{q}}$ and redefining expansion coefficients $\nu=N/p_t$, $\xi=\sqrt{p_q/p_t}$, and $a=p_\omega/p_t$ we can cast the pair propagator to a standard form that resembles the weak coupling result quoted earlier in Eq. \eqref{eq:L-approx} 
\begin{equation}\label{eq:L-Eliashberg}
L^R_{\bm{q}}(-i\Omega)=-\frac{1}{\nu}\frac{1}{\xi^2 q^2+t-ia\Omega}.
\end{equation}  
Further progress in estimating the effective coherence length $\xi$, damping parameter $a$, and current vertex $\bm{B}^{(e)}_{\bm{q}}$ can be made by utilizing a large parameter approximation. In the limit $\lambda\gg1$ Eliashberg interaction function in Eq. \eqref{eq:Lambda} can be approximated as $\Lambda\sim\lambda\omega^2_D/\varepsilon^2_n$, which translates to the self-energy $\Sigma^R\sim \lambda T_c$. Then carrying out the asymptotic analysis we estimate 
\begin{equation}\label{eq:estimates}
a\sim 1/(\lambda T_c), \quad \xi\sim(v_F/T_c)(1/\lambda^{3/2}), \quad \bm{B}^{(e)}_{\bm{q}}\sim e\nu\bm{q}(v_F/\lambda T_c)^2. 
\end{equation}
The principal observation to make here is that strong coupling effects change the relationship between the electrical current vertex and coherence length due to retardation of interaction mediated by phonons. Indeed, at weak coupling we see from Eq. \eqref{eq:Be-Ttau} in the ballistic limit $T\tau\gg1$ that $\bm{B}^{(e)}_{\bm{q}}\simeq e\nu (v_F/T_c)^2\bm{q} \propto \xi^2$ but the relationship $\bm{B}_{\bm{q}}\propto \xi^2$ does not hold at strong coupling in power counting of interaction parameter as now one finds $\bm{B}^{(e)}_{\bm{q}}\propto \lambda \xi^2$. 

At strong coupling fluctuation-induced correction to diagonal conductivity can be estimated with the help of the usual formula for AL correction but with the modified form of the pair propagator Eq. \eqref{eq:L-Eliashberg} and current vertex $\eqref{eq:estimates}$. Indeed, the electromagnetic response function is given by  
\begin{equation}
Q^{\text{AL}}_{xx}(\omega_\nu)=T\int_{\bm{q}}\sum_{\Omega_k}\big[\bm{B}^{(e)}_{\bm{q}}\big]^2_xL_{\bm{q}}(\Omega_k+\omega_\nu)L_{\bm{q}}(\Omega_k)
\end{equation}
which translates into the corresponding dc-conductivity  
\begin{equation}
\sigma^{\text{AL}}_{xx}=\frac{1}{4\pi T}\int_{\bm{q}}\big[\bm{B}^{(e)}_{\bm{q}}\big]^2_x\int^{+\infty}_{-\infty}\frac{[\Im L^R_{\bm{q}}(-i\Omega)]^2d\Omega}{\sinh^2(\Omega/2T)}=\frac{aT_c}{2\nu^2}\int_{\bm{q}}\big[\bm{B}^{(e)}_{\bm{q}}\big]^2_x\frac{1}{\big[\xi^2q^2+t\big]^3}
\end{equation}
after the standard steps of frequency summation and analytical continuation. As in all the previous similar cases, when performing frequency integral we expanded hyperbolic sine function in the denominator since integral is dominated by small values of frequency $\Omega\sim T-T_c$. The resulting temperature dependence of the AL term is the same as found in the BCS theory at weak coupling, 
\begin{equation}
\sigma^{\text{AL}}_{xx}\simeq \sigma_Q\lambda\left(\frac{T_c}{T-T_c}\right),
\end{equation} 
however it contains now an extra interaction parameter $\lambda\gg1$. This analysis can be extended to the calculation of fluctuation effects in the other relevant models of pairing boson. The method of extracting pair propagator for superconducting fluctuations near $T_c$ does not rely on any particular properties of the boson Green's function. The approximation in Eq. \eqref{eq:Gamma-approx} is based on the separation of energy scales between fermionic and bosonic modes which is generically valid at sufficient proximity to $T_c$. We plan to investigate strong coupling effects on other kinetic coefficients, including transversal anomalous responses, and collective modes. 

\begin{table}
\begin{center}
 \begin{tabular}{|c |c |c|c|}
 \hline
 $\delta\sigma^{\text{AHE}}_{xy}/\sigma_D$ & 2D limit & 3D limit & Eq. (\#)\\ 
 \hline
 Skew-scattering DOS & $\text{Gi}(\tau/\tau_{\text{sk}}) \ln(T_c\tgl)$ & $\sqrt{\text{Gi}} (\tau/\tau_{\text{sk}}) 1/\sqrt{T_c\tau_{\text{GL}}}$ & Eq. \eqref{sk-DOS-total}\\
 \hline
 Skew-scattering MT & $\text{Gi}(\tau/\tau_{\text{sk}})(T_c\tau_{\text{GL}})\ln(\tau_\phi/\tau_{\text{GL}})$ & $\sqrt{\text{Gi}} (\tau/\tau_{\text{sk}}) \sqrt{T_c\tau_{\text{GL}}}$ & Eq. \eqref{eq:sigma-sk-MT} \\
 \hline
 Nonlinear skew DOS & $\text{Gi}^2(\tau/\tau_{\text{sk}})(T_c\tau_{\text{GL}})^3$ & $\text{Gi}(\tau/\tau_{\text{sk}}) (T_c\tau_{\text{GL}})^2$ & Eq. \eqref{eq:sigma-sk-DOS-nl} \\
 \hline
 Interference skew MT & $\text{Gi}^2(\tau/\tau_{\text{sk}})(T_c\tau_{\text{GL}})^3\ln^2(\tau_\phi/\tau_{\text{GL}})$ & $\text{Gi}(\tau/\tau_{\text{sk}}) (T_c\tau_{\text{GL}})^2$ & Eq. \eqref{eq:sigma-sk-MT-nl} \\
 \hline
 Quantum-crossing AL & $\text{Gi}^2 (\tau/\tau_{\text{sk}}) (T_c\tau_{\text{GL}})\ln(T_c\tau_{\text{GL}})$ & $\text{Gi}(\tau/\tau_{\text{sk}})\sqrt{T_c\tau_{\text{GL}}}$ & Eq. \eqref{eq:sigma-sk-AL} \\
 \hline
 \end{tabular}
 \caption{Summary of results for the temperature dependence of the fluctuation-induced corrections to the anomalous Hall conductivity $\delta\sigma^{\text{AHE}}_{xy}$ in the main skew-scattering mechanism. The table entrees are normalized in units of normal state Drude conductivity $\sigma_D$. Ginzburg number has different definitions for 2D case $\text{Gi}\simeq1/(\nu T_c\xi^2)$, and for 3D case $\text{Gi}\simeq1/(\nu T_c\xi^3)^2$. }\label{table}
 \end{center}
 \end{table}

\section{Summary and outlook}\label{sec:summary}

In this work we presented a systematic approach to the anomalous Hall effect in superconductors. We explored the interplay of disorder scattering and electron interactions on the temperature dependence of anomalous Hall conductivity. We found that superconducting fluctuations close to $T_c$ contribute to transversal responses in both skew-scattering and side-jump mechanisms. We conclude that skew-scattering mechanism dominates as it leads to a more pronounced temperature dependence and stronger dependence on impurity scattering parameters. Table \ref{table} summarizes our main results for all the skew-scattering terms in density of states, Maki-Thompson, and Aslamazov-Larkin fluctuational effects for superconducting films (2D) and bulk (3D) systems.   

There remain a number of physically motivated interesting open questions that deserve further detailed studies. As an outlook for future possible research, we briefly discuss several effects and transport regimes that were left outside of the scope of this paper. 

(i) Spin-polarization plays an important role not only in defining skew-scattering cross-section but also in defining the structure of the pair-propagator \cite{Kee-Aleiner}. If the energy scale of spin-splitting becomes comparable or larger than temperature then fluctuational effects are governed by the virtual quasiparticle excitations that give dominant contributions to the triangular current vertex of AL diagram (virtual Cooper pairs). This regime has been analyzed in the context of longitudinal transport properties of Pauli limited superconductors \cite{KLC} and it is expected that anomalous responses will also have substantially different temperature dependences than that considered in this study. 

(ii) In many classes of unconventional superconductors $T_c$ can be suppressed to zero by a control parameter (e.g. doping). The effect of quantum superconducting fluctuations on anomalous Hall like responses near such a quantum critical point have not been systematically studied which is a conceptually interesting problem. 

(iii) We have concentrated primarily on the extrinsic mechanisms of AHE. However, it has been argued recently that the nontrivial band geometry of pairing electrons makes fingerprints at the spectrum of fluctuations \cite{Efimkin}. Indeed, under certain physically accessible conditions the spectrum of fluctuating Cooper pairs, as described by the effective Ginzburg-Landau Hamiltonian, can have topologically nontrivial Berry texture and is thus characterized by nonzero Chern number. As a result, such topological fluctuating pairs define an intrinsic anomalous Hall paraconductivity. The temperature dependence of this mechanism is weaker than that due to skew-scattering, however the idea itself is fruitful and may be more relevant in the other physical scenarios. 

(iv) Perhaps a more practical problem, is to extend current analysis to the case of strong impurities as it was done for example in Ref. \cite{Muttalib} for the normal state properties. This can be done by using a self-consistent expressions for the Green's function together with the Lippmann–Schwinger equation for the full $\hat{T}$-matrix. In the context of near-$T_c$ transport properties, we expect to obtain the same results for $\sigma^{\text{AHE}}_{xy}(T)$ in terms of singularities in the temperature dependence but with properly renormalized scattering times. However, below $T_c$ full $\hat{T}$-matrix analysis may lead to new features such as formation of sub-gap bands in unconventional superconductors that could have significant effect on the observed anomalies in the polar Kerr effect.     

\section{Acknowledgments}

We thank Andrey Chubukov, Maxim Dzero, Yuxuan Wang, and Emil Yuzbashyan for valuable discussions on the broad range of topics related to Eliashberg strong-coupling theory.  
We also thank Yuli Lyanda-Geller and Pavel Ostrovsky for the extremely helpful correspondence regarding the Hikami box calculation and for bringing Ref. \cite{Aronov-Hikami} to our attention. This work at UW-Madison was supported by NSF CAREER Grant No. DMR-1653661 and the Ray MacDonald Endowment Award. 
This work was performed in part at Aspen Center for Physics, which is supported by National Science Foundation Grant PHY-1607611.

\end{document}